\documentclass[%
twocolumn,
superscriptaddress,
longbibliography
nobibnotes,
amsmath,amssymb,
aps,
pre,
]{revtex4-2}

\usepackage{graphicx}
\usepackage{dcolumn}
\usepackage[hypertexnames=false]{hyperref}
\usepackage{txfonts}
\usepackage{tabularx}
\newcolumntype{C}{>{\centering\arraybackslash}X}
\newcolumntype{L}{>{\raggedright\arraybackslash}X}
\newcolumntype{R}{>{\raggedleft\arraybackslash}X}
\usepackage{multirow}
\usepackage{bm}
\usepackage{braket}

\hypersetup{
    colorlinks=true,
    linkcolor=blue,
    citecolor=blue,
    urlcolor=blue,
}

\usepackage{color}
\newcommand{\red}[1]{{\color{black} #1}}

\begin{document}
\title{Universal properties of repulsive self-propelled particles and attractive driven particles}
\author{Hiroyoshi Nakano}
\affiliation{Institute for Solid State Physics, University of Tokyo, 5-1-5, Kashiwanoha, Kashiwa 277-8581, Japan}
\author{Kyosuke Adachi}
\affiliation{Nonequilibrium Physics of Living Matter RIKEN Hakubi Research Team, RIKEN Center for Biosystems Dynamics Research (BDR), 2-2-3 Minatojima-minamimachi, Chuo-ku, Kobe 650-0047, Japan}
\affiliation{RIKEN Interdisciplinary Theoretical and Mathematical Sciences Program (iTHEMS), 2-1 Hirosawa, Wako 351-0198, Japan}

\date{\today}
\begin{abstract}
Motility-induced phase separation (MIPS) is a nonequilibrium phase separation that has a different origin from equilibrium phase separation induced by attractive interactions.
Similarities and differences in collective behaviors between these two types of phase separation have been intensely discussed.
Here, to study another kind of similarity between MIPS and attraction-induced phase separation under a nonequilibrium condition, we perform simulations of active Brownian particles with uniaxially anisotropic self-propulsion (uniaxial ABPs) in two dimensions.
We find that (i) long-range density correlation appears in the homogeneous state, (ii) anisotropic particle configuration appears in MIPS, where the anisotropy removes the possibility of microphase separation suggested for isotropic ABPs [X.-Q. Shi \textit{et al}., Phys. Rev. Lett. 125, 168001 (2020)], and (iii) critical phenomena for the anisotropic MIPS presumably belong to the universality class for two-dimensional uniaxial ferromagnets with dipolar long-range interactions.
Properties (i)-(iii) are common to the well-studied randomly driven lattice gas (RDLG), which is a particle model that undergoes phase separation by attractive interactions under external driving forces, suggesting that the origin of phase separation is not essential for macroscopic behaviors of uniaxial ABPs and RDLG.
Based on the observations in uniaxial ABPs, we construct a coarse-grained Langevin model, which shows properties (i)-(iii) and corroborates the generality of the findings.
\end{abstract}
\maketitle

\section{Introduction}
\label{sec:introduction}
Liquid-gas or liquid-liquid phase separation is a typical collective phenomenon that has been observed in a wide range of systems from polymer solution~\cite{Rubinstein2003} to biological materials~\cite{Banani2017,Berry2018}.
Basically, equilibrium phase separation is caused by attractive interactions between molecules or particles~\cite{Rubinstein2003}, and the corresponding critical phenomena have been considered to belong to the Ising universality class~\cite{Guggenheim1945,Kadanoff1967,Watanabe2012}.
In contrast, in nonequilibrium systems, depending on how the detailed balance is broken, the critical exponents for phase separation can deviate from the Ising model values~\cite{Schmittmann1995}, and phase separation can emerge from different mechanisms such as chemical reactions~\cite{Cotton2022} and coupling to multiple heat baths~\cite{Weber2016}.
A comprehensive understanding of the seemingly broad spectrum of nonequilibrium phase separation requires theoretical studies from a unified viewpoint.

For attractively interacting particles that undergo phase separation, one of the ways to break the detailed balance is external driving with bulk fields or boundary reservoirs, which generically changes the density correlation to long-ranged~\cite{Ronis1982,Garrido1990,Derrida1993,Dorfman1994,Schmittmann1995,De_Zarate2006,Takacs2011,Nakano2022} and leads to nonequilibrium critical phenomena~\cite{Onuki1979,Katz1984,Schmittmann1995,Chakrabarti1999,Tauber2002,Haga2015,Nakano2021}.
The driven lattice gas (DLG)~\cite{Katz1984,Marro1999} and randomly driven lattice gas (RDLG)~\cite{Garrido1990,Cheng1991,Hwang1993,Praestgaard1994,Praestgaard2000,Albano2002,Basu2017,Volpati2017} are prototypical models of nonequilibrium phase separation, in which particles stochastically move with short-range attractive interactions under external driving forces.
Unidirectional and uniaxial driving forces are assumed in DLG and RDLG, respectively, which makes the difference in symmetry between these two models.
In DLG and RDLG, spatial anisotropy of the driving force causes long-range density correlation~\cite{Hwang1993} and critical phenomena that do not belong to the Ising universality class~\cite{Schmittmann1991,Schmittmann1993,Caracciolo2005,Li2019}.
In particular, the universality class for RDLG has been considered as that for uniaxial ferromagnets with dipolar long-range interactions~\cite{Aharony1973,Brezin1976}, according to the renormalization group (RG) analysis~\cite{Schmittmann1993,Praestgaard2000}.

% ---------------- figure ----------------
\begin{figure*}[t]
\centering
\includegraphics[scale=1.0]{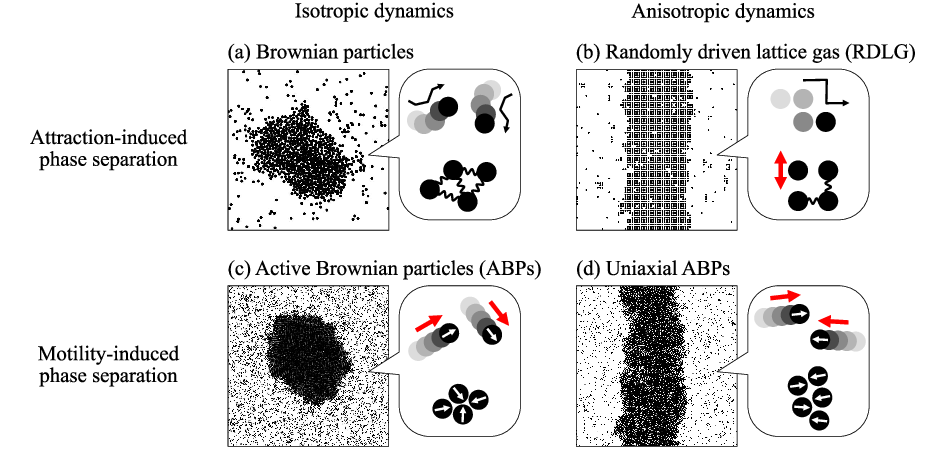}
\caption{Four types of phase separation.
The row and column correspond to the type of phase separation (attraction- or motility-induced) and the type of dynamics (isotropic or anisotropic), respectively.
In each panel, a typical particle configuration obtained from model simulations is shown with schematic figures of the single particle motion and small cluster formation.
(a) Brownian particles follow overdamped dynamics with attractive interactions (wavy lines) and random forces.
(b) In RDLG, particles stochastically move with attractive interactions (wavy lines) and external driving force (red arrow) along an axis (i.e., $y$-axis in the figure).
(c) ABPs show self-propelled motion (red arrow) with repulsive interactions and random forces.
(d) Uniaxial ABPs show anisotropic self-propelled motion favored along an axis (i.e., $x$-axis in the figure) with repulsive interactions and random forces similar to (c) [see Eq.~\eqref{Eq:ABP} for the detail].
}
\label{fig_schematic}
\end{figure*}
% ----------------------------------------

Self-propulsion is another way to break the detailed balance~\cite{Chate2020,Gompper2020,Bowick2022}.
In a crowd of self-propelled particles, or active matter, collective phenomena ranging from giant number fluctuations~\cite{Toner2012,Shankar2018} to active turbulence~\cite{Dunkel2013} have been found using biological~\cite{Szabo2006,Nishiguchi2017,Kawaguchi2017,Liu2019,Tan2022} and artificial~\cite{Deseigne2010,Palacci2013,Bricard2013,Kumar2014,Ginot2018,Deblais2018,Geyer2019,Chardac2021,Wang2021} systems.
In particular, as shown in simulations~\cite{Peruani2011,Thompson2011,Fily2012,Fodor2016,Whitelam2018} and experiments~\cite{Buttinoni2013}, self-propelled particles with repulsive interactions can undergo phase separation, which is called motility-induced phase separation (MIPS)~\cite{Cates2015}.
No attractive interactions are necessary for MIPS, which is distinct from equilibrium phase separation or nonequilibrium phase separation under external driving.
MIPS has been studied in comparison with equilibrium phase separation, and similarities and differences between them have been reported [see Figs.~\ref{fig_schematic}(a) and (c)].
For example, the global phase diagrams for MIPS~\cite{Digregorio2018,Omar2021} and equilibrium phase separation are similar if we exchange the axis of self-propulsion strength for MIPS with that of attractive interaction strength for equilibrium phase separation.
In addition, the lever rule~\cite{Rubinstein2003}, which is common to equilibrium phase separation, holds for MIPS in particle models~\cite{Shi2020}, and consistently, effective free energy has been proposed based on coarse-grained models~\cite{Solon2018a,Solon2018b}.
In contrast, it is still unclear whether the critical phenomena for MIPS belong to the Ising universality class~\cite{Partridge2019,Maggi2021,Dittrich2021,Speck2022}.
Furthermore, as a unique feature of MIPS, the nucleation of persistent gas bubbles that can lead to microphase separation has been found~\cite{Tjhung2018,Singh2019,Caporusso2020,Shi2020}.

In the previous work~\cite{Adachi2022}, one of the authors has proposed another kind of similarity between the anisotropic version of MIPS and attraction-induced phase separation under external driving.
Briefly, it has been found that a lattice gas model with spatially anisotropic self-propulsion exhibits a variety of collective behaviors: long-range density correlation, anisotropic phase separation, and critical phenomena with the universality class expected to be the same as that for uniaxial dipolar ferromagnets.
All these behaviors have also been seen in RDLG, which indicates a connection between repulsively interacting particles with anisotropic self-propulsion and attractively interacting particles under external driving.
However, the generality of such observations is still unclear beyond the considered lattice gas model.
In particular, though persistent gas bubbles have been observed in active Brownian particles (ABPs)~\cite{Shi2020}, a prototypical model of MIPS~\cite{Fily2012}, the fate of gas bubbles under spatial anisotropy has not been investigated.
More broadly, systematic studies of the effect of spatial anisotropy on active matter are still scarce~\cite{Mishra2014,Brambati2022,Solon2022,Chatterjee2022,Broker2022}.

In this paper, toward a comprehensive understanding of the relation between the anisotropic MIPS and attraction-induced phase separation under external driving, we consider ABPs with anisotropic self-propulsion.
In Fig.~\ref{fig_schematic}, we show typical particle configurations obtained from model simulations for the above-mentioned four types of phase separation: attraction/motility-induced phase separation with isotropic/anisotropic dynamics.
In each panel of Fig.~\ref{fig_schematic}, we also schematically show the single particle motion and typical configuration of small clusters, which can grow up to a macroscopic scale and lead to phase separation.
Our present focus is on the relation between the two types of anisotropic phase separation in the right panels of Fig.~\ref{fig_schematic}.

We perform simulations of ABPs with uniaxially anisotropic self-propulsion (uniaxial ABPs).
We find that, as expected from the previous study~\cite{Adachi2022}, uniaxial anisotropy dramatically changes the collective behaviors and causes long-range correlation, anisotropic phase separation, and critical phenomena that are presumably in the same universality class as that for uniaxial dipolar ferromagnets.
Furthermore, uniaxial anisotropy suppresses the growth of gas bubbles in MIPS~\cite{Shi2020} and stabilizes macroscopic phase separation.
Developing a coarse-grained model for particles with anisotropic self-propulsion, we corroborate the generality of the observed phenomena.

% ---------------- figure ----------------
\begin{figure*}[t]
\centering
\includegraphics[scale=1]{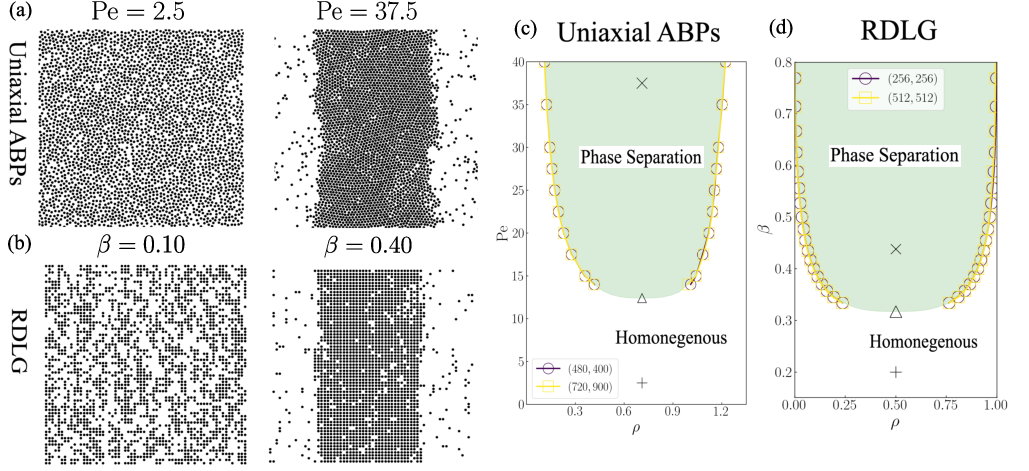}
\caption{Phase behaviors of uniaxial ABPs and RDLG.
(a) Typical snapshots of uniaxial ABPs.
The parameters are $(L_x,L_y)=(64,64)$, $\rho=0.71$ ($N=2908$), $(\mu_{\parallel},\mu_{\perp},\mu_{\theta})=(1,0.25,\red{1.5})$, $\epsilon = 0.01$, $\mathrm{Pe} = 2.5$ (left), and $\mathrm{Pe} = 37.5$ (right).
(b) Typical snapshots of RDLG.
The parameters are $(L_x,L_y)=(64,64)$, $\rho=0.5$ ($N=2048$), $E=100$, $\beta=0.1$ (left), and $\beta=0.40$ (right).
(c) Phase diagram of uniaxial ABPs.
The parameters are the same as those for (a).
(d) Phase diagram of RDLG. The parameters are the same as those for (b).
In (c) and (d), the signs (+/$\times$) indicate the parameter sets where the left/right panels of (a) and (b) are calculated, respectively.
The triangle ($\triangle$) represents the estimated critical point, the properties of which are discussed in Sec.~\ref{sec:connection_to_uniaxial_dipolar_ferromagnets}.}
\label{fig_micro_phasediagram}
\end{figure*}
% ----------------------------------------

\section{Microscopic models}
\label{sec:microscopic_models}
In this section, we explain the numerical implementation of uniaxial ABPs and RDLG, which are anisotropic extensions of the isotropic ABPs and equilibrium lattice gas, respectively.
We also present phase diagrams for the two models, which provide preliminary insights into collective behaviors.

\subsection{Active Brownian particles with uniaxial anisotropy}
For uniaxial ABPs, $N$ particles are confined in $[0,L_x] \times [0,L_y]$ with periodic boundary conditions. The state of the $i$th particle is specified by position $\bm{r}_i$ and polarity angle $\theta_i$.
The time evolution of ($\bm{r}_i$, $\theta_i$) is governed by
\begin{equation}
\left\{
\begin{array}{l}
     \displaystyle \frac{d r^a_i}{dt} = \mu_i^{ab} \left[ F_{0} n_i^b  -\sum_{j (\neq i)} \frac{\partial V(|\bm{r}_i-\bm{r}_j|)}{\partial r^b_i} \right] + \eta_i^{a} \\
    \displaystyle \frac{d \theta_i}{dt} = - \epsilon \frac{\partial U(\theta_i)}{\partial \theta_i} + \sqrt{2\tau \mu_{\theta}}\xi^{\theta}_i,
\end{array}
\label{Eq:ABP}
\right.
\end{equation}
where $a \in \{ x, y \}$, $\mu_i^{ab} := \mu_{\parallel} n_i^a n_i^b + \mu_{\perp}(\delta^{ab} - n_i^a n_i^b)$, $\eta_i^{a} := \sqrt{2\tau \mu_{\parallel}} \xi_i^{\parallel} n_i^a + \sqrt{2\tau \mu_{\perp}} \xi_i^{\perp} (\bm{n}_i\times \hat{\bm{z}})^a$, and $\bm{n}_i := (\cos \theta_i, \sin \theta_i)$.
Also, $\xi_i^{\parallel}$, $\xi_i^{\perp}$, and $\xi_i^{\theta}$ are Gaussian white noises with zero mean and unit variance. \red{The translational noise $\eta_i^a$, representing thermal noise, is added to satisfy the detailed-balance condition when the self-propulsion force $F_0 n_i^b$ is absent.} We assume the two-body interaction as $V(r)=(k/2)(\sigma-r)^2$ for $r < \sigma$ and $V(r)=0$ otherwise.
The potential for the polarity angle, $U (\theta)$, is added to model the effect of spatial anisotropy on self-propulsion, and $\epsilon$ ($\geq 0$) represents the strength of anisotropy.
In this work, we use a simple potential function, $U(\theta) = -\cos(2\theta)$, which enhances the alignment of polarity along the $x$-axis (i.e., $\theta=0$ or $\pi$).
Note that the polarity angle of each particle can take any value between $0$ and $2 \pi$, in contrast to the previous model~\cite{Adachi2022}, in which the polarity angle is restricted to $0$ or $\pi$.
We also stress that we consider anisotropy of the self-propulsion direction, not of the particle shape.

\red{
In the case of $\epsilon = 0$, the properties of the model [Eq.~(\ref{Eq:ABP})] have been studied in Ref.~\cite{Shi2020}. In particular, the anisotropic mobility tensor $\mu_i^{ab}$ has been used to enhance the nucleation of gas bubbles in the phase-separated state. In our numerical simulations, we follow Ref.~\cite{Shi2020} and use anisotropic $\mu_i^{ab}$ for the case of $\epsilon > 0$. While the anisotropy of $\mu_i^{ab}$ results in the polarity-dependent response of particle motion to the force, it does not induce spatially anisotropic particle motion along a fixed axis, in contrast to the spatial anisotropy caused by $\epsilon$. Thus, in the following, we use the term ``anisotropy'' to refer to the effects of $\epsilon$.
}

Throughout the numerical study, we set $\sigma=1$, $k=20$, and $\tau=0.01$. The controlled parameters are system size $(L_x,L_y)$, particle density $\rho:=N/(L_x L_y)$, mobilities $(\mu_{\parallel},\mu_{\perp},\mu_{\theta})$, magnitude of anisotropy $\epsilon$, and the P{\'e}clet number, $\textrm{Pe}:=F_0\sigma/\tau$, which represents the dimensionless strength of self-propulsion. The simulations are performed using LAMMPS~\cite{Plimpton1995,Thompson2022}. The time integration is performed by the Euler method with timestep $dt = 0.02$.

Figure~\ref{fig_micro_phasediagram}(a) displays snapshots with two sets of parameters, which show that this model undergoes anisotropic phase separation.
We stress that there is no attractive interaction in uniaxial ABPs, just like isotropic ABPs. As suggested in Fig.~\ref{fig_schematic}(d), this phase separation originates from the self-propulsion of each particle. We also present the phase diagram in Fig.~\ref{fig_micro_phasediagram}(c); phase separation emerges for large $\textrm{Pe}$, which is also the same as in isotropic ABPs. Thus, this phase separation is regarded as the anisotropic extension of isotropic MIPS [see Fig.~\ref{fig_schematic}(c)].

\subsection{Randomly driven lattice gas}
For RDLG, we consider $N$ particles on a square lattice with system size $(L_x,L_y)$ in units of the lattice constant. The state of the $i$th site is specified by occupation number $n_i$, and the set of $n_i$ represents the configuration of the whole system. We assume exclusion between particles so that each site can be occupied by at most one particle, i.e., $n_i \in \{0,1\}$. We also consider attractive interaction between neighboring particles, which is represented by the following Hamiltonian:
\begin{eqnarray}
H = - J \sum_{\langle i,j\rangle} n_{i} n_{j}.
\end{eqnarray}

The state of the system is updated in three steps:
\begin{enumerate}
    \item We randomly choose two adjacent sites, $(i,j)$, and calculate the energy difference ($\Delta H$) between the original configuration and the new configuration obtained by exchanging the state of the $i$th site with the state of the $j$th site.
    \item If sites $(i,j)$ are located along the $x$-axis, the new configuration is accepted with probability $\min(1,e^{-\beta \Delta H})$.
    \item If sites $(i,j)$ are located along the $y$-axis, the new configuration is accepted with probability $\min(1,e^{-\beta (\Delta H + E \eta)})$, where $E$ is the strength of the \red{driving force}, and $\eta$ is a random number drawn from a Gaussian distribution with zero mean and unit variance.
\end{enumerate}

For step 3, the random driving force is applied along the $y$-axis. We basically set the parameters to $J=4$ and $E=100$ and control $\beta$ and $\rho := N/L_xL_y$.
\red{
Note that our numerical implementation leads to the same type of macroscopic behaviors as those in the previous studies~\cite{Praestgaard2000, Achahbar2001}.
}

\red{
Since the driving force along the $y$-axis (i.e., $E \eta$) competes with and effectively weakens the attractive interaction (i.e., $\Delta H$), the motion of particles that interact with the neighbors is enhanced along the $y$-axis.
In particular, $E=100$ is practically equivalent to the limiting case with $E = \infty$, where the configuration is updated regardless of the value of $\Delta H$.
This limiting case has been commonly used in simulations of DLG and RDLG~\cite{Schmittmann1995}.
}

Figure~\ref{fig_micro_phasediagram}(b) displays snapshots with two sets of parameters. 
This model undergoes phase separation induced by the attractive interaction [Fig.~\ref{fig_schematic}(b)] though the motion of each particle is affected by the random driving force. We present the phase diagram in Fig.~\ref{fig_micro_phasediagram}(d); phase separation is controlled by inverse temperature $\beta$, just like in equilibrium particle systems with attractive interactions [see Fig.~\ref{fig_schematic}(a)].

\subsection{Orientation of phase separation}
\label{sec:global_phase_diagram}
Self-propulsion is favored along the $x$-axis in uniaxial ABPs, while the driving force is applied along the $y$-axis in RDLG.
Despite this difference in the direction of the enhanced particle motion, the dense and dilute regions are segregated along the $x$-axis in both uniaxial ABPs and RDLG [see Figs.~\ref{fig_micro_phasediagram}(a) and (b)].
Such a coincidence of the collective behavior can be interpreted from a microscopic viewpoint as follows.
For uniaxial ABPs, self-propulsion induces persistent collision of particles along the $x$-axis, leading to effective adhesion between particles along the $x$-axis.
Since this type of collision is less probable along the $y$-axis, particles can move more freely along the $y$-axis.
Thus, particle clusters that are caused by the effective adhesion should be elongated along the $y$-axis, which results in the segregation along the $x$-axis [see Fig.~\ref{fig_schematic}(d)].
Note that similar cluster patterns have been recently found in simulations of ABPs with anisotropic self-propulsion~\cite{Broker2022}.
For RDLG, the driving force enhances the free motion of particles along the $y$-axis.
Thus, particle clusters caused by the attractive interaction should be elongated in the $y$ direction, leading to the segregation along the $x$-axis [see Fig.~\ref{fig_schematic}(b)].
See Appendix~\ref{App:diffandsimi_micro} for further comparisons between uniaxial ABPs and RDLG.

% ---------------- figure ----------------
\begin{figure}[bt]
\centering
\begin{center}
\includegraphics[scale=1.0]{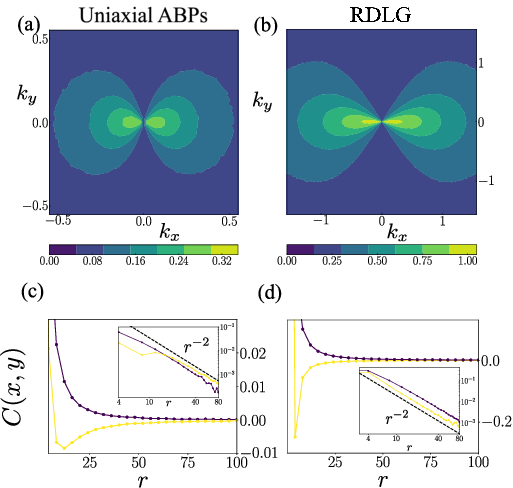} 
\end{center}
\caption{Singular structure factors in the homogeneous states.
(a, b) Heatmap of structure factor $S(\bm{k})$ for (a) uniaxial ABPs and (b) RDLG.
(c, d) Density correlation functions $C(x,0)$ (yellow) and $C(0,y)$ (purple) for (c) uniaxial ABPs and (d) RDLG.
In the insets, the absolute value is plotted on the log-log scale.
The parameters used for (a, c) and (b, d) are the same as those for the left panels of Figs.~\ref{fig_micro_phasediagram}(a) and (b), respectively.
The system size is set to $L_x=L_y=360$ for both models.
}
\label{fig_micro_homobehavior}
\end{figure}

\begin{figure}[bt]
\centering
\begin{center}
\includegraphics[scale=1.0]{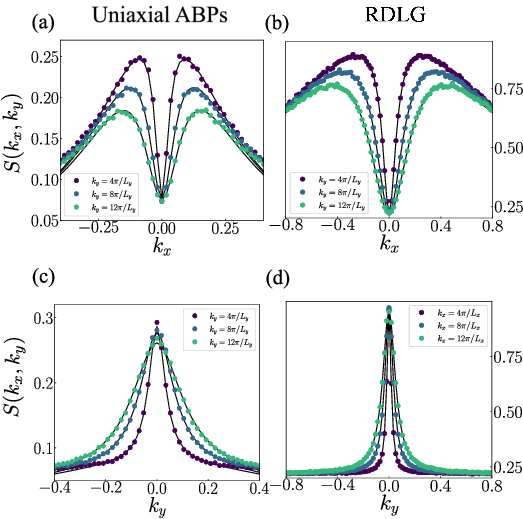} 
\end{center}
\caption{Quantitative comparison between the simulated structure factor and the theoretical expression [Eq.~\eqref{Eq:strfac_linear}], where the same data as plotted in Fig.~\ref{fig_micro_homobehavior} is used.
(a, b) Structure factor $S(k_x,k_y)$ with $k_y=4\pi/L_y$, $8\pi/L_y$, and $12\pi/L_y$ for (a) uniaxial ABPs and (b) RDLG.
(c, d) Structure factor $S(k_x,k_y)$ with $k_x=4\pi/L_x$, $8\pi/L_x$, and $12\pi/L_x$ for (c) uniaxial ABPs and (d) RDLG.
In all figures, the colored dots represent the simulation results, and the black lines represent the theoretical expression with the best-fit parameter.}
\label{fig_micro_homobehavior2}
\end{figure}
% ----------------------------------------

\section{Properties of homogeneous state}
\label{sec:homogeneous_state_properties}
Hydrodynamic descriptions are helpful in understanding the collective behavior of particles.
For RDLG, homogeneous state properties have been studied using a linear coarse-grained model~\cite{Schmittmann1998,Praestgaard2000}:
\begin{equation}
\partial_t \phi + \bm{\nabla} \cdot \bm{j} = 0
\label{Eq:cgmodel_linear}
\end{equation}
with
\begin{equation}
\left\{
\begin{array}{l}
     \displaystyle j_x = - \partial_x (a_x -K_{xx}\partial_x^2 - K_{xy}\partial_y^2)\phi + \sqrt{2D_x} \xi_x \\
    \displaystyle j_y = - \partial_y (a_y -K_{yx}\partial_x^2 - K_{yy}\partial_y^2)\phi + \sqrt{2D_y} \xi_y. \nonumber
\end{array}
\right.
\end{equation}
Here, $\phi (\bm{r}, t)$ is the density fluctuation field, $\bm{\xi} (\bm{r}, t)$ is a Gaussian noise with $\braket{\xi_a (\bm{r}, t)} = 0$ and $\braket{\xi_a (\bm{r}, t) \xi_b (\bm{r'}, t')} = \delta_{ab} \delta (\bm{r} - \bm{r}') \delta (t - t')$.
In the isotropic limit ($K_{xx}=K_{xy}=K_{yx}=K_{yy}=K$, $a_x=a_y=a$, and $D_x=D_y=D$), Eq.~\eqref{Eq:cgmodel_linear} is reduced to the so-called model B~\cite{Hohenberg1977},
\begin{equation}
\partial_t \phi = \bm{\nabla}^2 \frac{\delta \mathcal{H}}{\delta \phi} - \sqrt{2D} \bm{\nabla} \cdot \bm{\xi},
\label{Eq:cgmodel_linear_isotropic_1}
\end{equation}
where $\mathcal{H}$ is a coarse-grained Hamiltonian:
\begin{equation}
\mathcal{H} = \int d^2\bm{r} \left[ \frac{a}{2}\phi^2 + \frac{K}{2}(\bm{\nabla} \phi)^2 \right].
\label{Eq:cgmodel_linear_isotropic_2}
\end{equation}
Thus, Eq.~\eqref{Eq:cgmodel_linear} is regarded as an extension of model B to an anisotropic system that respects the symmetry of particle dynamics in RDLG.

In the following, we demonstrate that the homogeneous states of uniaxial ABPs and RDLG exhibit the same type of long-range correlation as a generic feature of the nonequilibrium collective dynamics, which can be explained by Eq.~\eqref{Eq:cgmodel_linear}.
In Appendix~\ref{App:dipolar_ferromagnet}, using the well-known correspondence between RDLG and uniaxial dipolar ferromagnets~\cite{Schmittmann1993,Praestgaard2000}, we further establish the connection between uniaxial ABPs and dipolar ferromagnets.

\subsection{Long-range density correlation}
The steady-state long-range correlation of a conserved quantity has been recognized as a general feature of nonequilibrium systems with anisotropic dynamics~\cite{Garrido1990,Dorfman1994}.
Specifically, the fluctuation of a conserved quantity, which we denote as $\delta A(\bm{r})$ here, decays as
\begin{equation}
\braket{\delta A(\bm{r})\delta A(\bm{r}')} \sim c_{\rm eq} e^{-|\bm{r}-\bm{r}'|/\xi} + \frac{c_{\rm neq}}{|\bm{r}-\bm{r}'|^{\alpha}},
\label{Eq:long-range_correlation}
\end{equation}
where $\braket{\cdot}$ is an ensemble average in the steady state, and $c_{\rm eq}$ and $c_{\rm neq}$ are constants.
The first term represents an exponential decay that also appears in equilibrium systems, while the second term is a nonequilibrium correction that leads to the long-range correlation with a power-law decay.
The presence of long-range correlation (i.e., $c_\mathrm{neq} \neq 0$) is ubiquitous in nonequilibrium systems with spatial anisotropy.

In uniaxial ABPs and RDLG, the self-propulsion and driving force violate the detailed balance in a spatially anisotropic way, respectively.
Thus, the long-range correlation of the density field, which is a locally conserved field, is expected to appear in both systems.
Though RDLG has been known to show the long-range correlation~\cite{Garrido1990,Schmittmann1998,Praestgaard2000}, for completeness, we explain the results for uniaxial ABPs and RDLG in parallel.
Assuming small self-propulsion $\mathrm{Pe}$ in uniaxial ABPs and low inverse temperature $\beta$ in RDLG [see the plus sign (+) in Figs.~\ref{fig_micro_phasediagram}(c) and (d)], we focus on typical homogeneous states [Figs.~\ref{fig_micro_phasediagram}(a) and (b)].
We calculate the structure factor and the two-point correlation function, which are defined as
\begin{equation}
S(\bm{k}) := \frac{1}{L_x L_y}\big\langle |\delta \tilde{\rho}(\bm{k})|^2\big\rangle
\end{equation}
and
\begin{equation}
C(\bm{r}) := \big\langle \delta\rho(\bm{r}) \delta\rho(\bm{0})\big\rangle,
\end{equation}
respectively.
Here, $\rho(\bm{r}):= \sum_{i=1}^N \delta(\bm{r}-\bm{r}_i)$, $\delta\rho(\bm{r}) := \rho(\bm{r}) - \langle\rho(\bm{r})\rangle$, and $\tilde{\rho} (\bm{k})$ is the Fourier transformation of $\rho(\bm{r})$.

We show the heatmaps of $S(\bm{k})$ for uniaxial ABPs and RDLG in Figs.~\ref{fig_micro_homobehavior}(a) and (b), respectively, both of which exhibit owl-like or butterfly-like patterns~\cite{Schmittmann1998}.
Analytically, the observed pattern of $S(\bm{k})$ can be characterized by the discontinuity at the origin in the Fourier space, i.e.,
\begin{eqnarray}
\lim_{k_x \to 0} S(k_x, k_y = 0) \neq \lim_{k_y \to 0} S(k_x = 0, k_y).
\end{eqnarray}
This discontinuity of $S(\bm{k})$ reflects the power-law decay of $C(\bm{r})$ in the real space~\cite{Schmittmann1998}.
As shown in Figs.~\ref{fig_micro_homobehavior}(c) and (d), the correlation function [$C(x,y=0)$ (yellow) and $C(x=0,y)$ (purple)] indeed shows a power-law decay as $\sim r^{-2}$, which implies the long-range density correlation.
The negative correlation observed in $C(x, y = 0)$ suggests the formation of transient clusters elongated along the $y$-axis.
This orientation of clusters is consistent with the configurations in phase separation shown in Figs.~\ref{fig_micro_phasediagram}(a) and (b) (see Sec.~\ref{sec:global_phase_diagram}).

% ---------------- figure ----------------
\begin{figure*}[t]
\centering
\includegraphics[scale=1]{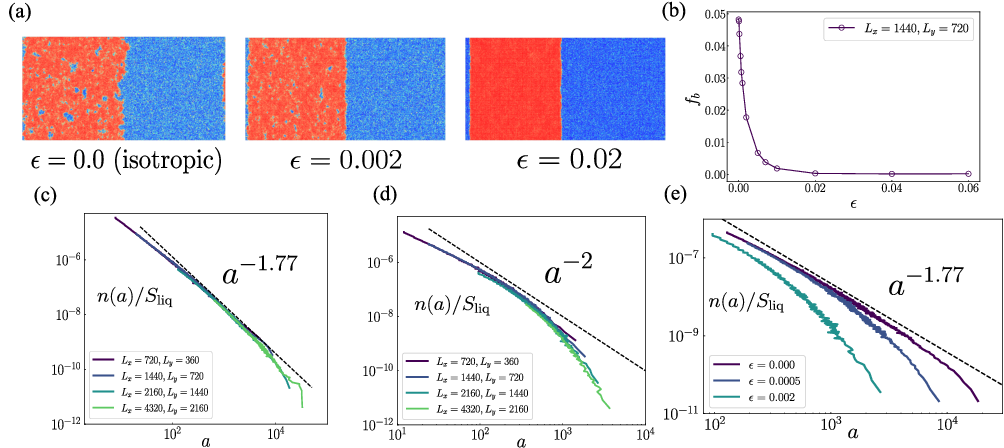}
\caption{Persistent gas bubbles in the phase-separated state of uniaxial ABPs.
(a) Typical snapshots in the steady state at $(L_x,L_y)=(1440,720)$ for three values of $\epsilon$.
The colors represent the particle density from $0$ (blue) to $1.5$ (red).
(b) Bubble fraction $f_b$ as a function of $\epsilon$ for $(L_x,L_y)=(1440,720)$.
(c, d) Bubble size distribution divided by the total liquid area, $n(a)/S_{\rm liq}$, for (c) isotropic ($\epsilon=0$) and (d) anisotropic ($\epsilon=0.002$) systems.
(e) $n(a)/S_{\rm liq}$ for three values of $\epsilon$ for $(L_x,L_y)=(2880,1440)$.
In all figures, the parameters are chosen as $\rho=0.765$, $(\mu_{\parallel},\mu_{\perp},\mu_{\theta})=(1,0.3125,2.75)$, and $\textrm{Pe} = 100$.}
\label{fig_micro_phasediagram_app}
\end{figure*}
% ----------------------------------------

\subsection{Linear coarse-grained model}
According to the previous studies, the owl-like pattern of the structure factor observed in RDLG [Fig.~\ref{fig_micro_homobehavior}(b)] can be reproduced by the linear coarse-grained model [Eq.~\eqref{Eq:cgmodel_linear}]~\cite{Schmittmann1998}.
The similar pattern observed in uniaxial ABPs [Fig.~\ref{fig_micro_homobehavior}(a)] suggests that uniaxial ABPs and RDLG share the same macroscopic dynamics described by Eq.~\eqref{Eq:cgmodel_linear}.
To confirm the validity of Eq.~\eqref{Eq:cgmodel_linear} for both uniaxial ABPs and RDLG, we examine the structure factor for the coarse-grained density fluctuation, $S_\mathrm{lin} (\bm{k}):= \braket{|\tilde{\phi} (\bm{k})|^2} / (L_x L_y)$, and $\tilde{\phi} (\bm{k})$ is the Fourier transformation of $\phi (\bm{r})$.
From Eq.~(\ref{Eq:cgmodel_linear}), we can obtain~\cite{Schmittmann1995,Schmittmann1998}
\begin{equation}
    S_\mathrm{lin} (\bm{k}) = \frac{D_x {k_x}^2 + D_y {k_y}^2}{a_x {k_x}^2 + a_y {k_y}^2 + K_{xx} {k_x}^4 + 2K_{xy} {k_x}^2{k_y}^2 + K_{yy} {k_y}^4}.
    \label{Eq:strfac_linear}
\end{equation}

For uniaxial ABPs, we fit the simulation data of $S(\bm{k})$ for $\bm{k}\in [2\pi/L_x,20\pi/L_x] \times [2\pi/L_y,20\pi/L_y]$ with Eq.~\eqref{Eq:strfac_linear}, using $D_x$, $D_y$, $a_x$, $a_y$, $K_{xy}$, and $K_{yy}$ as fitting paramters with $K_{xx} = 1$.
The fitting results are as follows:
\begin{align}
    & D_x = 0.0287, \ D_y = 0.00600,  a_x = 0.0990, \ a_y = 0.0778,\nonumber \\
    & K_{xy}=0.525,\ K_{yy} =0.145.
\end{align}
In Figs.~\ref{fig_micro_homobehavior2}(a) and (c), we plot the observed $S (\bm{k})$ (with dots) and the fitted $S_\mathrm{lin} (\bm{k})$ (with lines).
The results show that Eq.~\eqref{Eq:strfac_linear} quantitatively reproduces the observed behavior of the structure factor for small $|\bm{k}|$, which reflects the long-wavelength density fluctuation.
We also fit the simulation data of RDLG in the same way as used for uniaxial ABPs.
The fitting results are as follows:
\begin{align}
    & D_x = 1.37, \ D_y = 1.00, \ a_x = 1.41, \ a_y = 4.52,\nonumber \\
    & K_{xy}=0.609,\ K_{yy} = -0.0899.
\end{align}
In Figs.~\ref{fig_micro_homobehavior2}(b) and (d), we compare the observed $S (\bm{k})$ and the fitted $S_\mathrm{lin} (\bm{k})$, which show quantitative agreement as expected.

As discussed in previous studies of DLG and RDLG~\cite{Schmittmann1995}, we can derive the asymptotic behavior of the long-range part of the correlation function, $C_\mathrm{lin} (\bm{r})$, which is the inverse Fourier transformation of $S_\mathrm{lin} (\bm{k})$.
From Eq.~\eqref{Eq:strfac_linear} we can obtain
\begin{equation}
    C_\mathrm{lin} (x, 0) \sim -x^{-2}, \ C_\mathrm{lin} (0, y) \sim y^{-2} \ \ \ (r \to \infty),
\end{equation}
which is also consistent with the power-law decay of $C(\bm{r})$ observed in uniaxial ABPs [Fig.~\ref{fig_micro_homobehavior}(c)] and RDLG [Fig.~\ref{fig_micro_homobehavior}(d)].

\section{Phase separation properties}
\label{sec:phase_separation_properties}
As briefly explained in Sec.~\ref{sec:microscopic_models}, uniaxial ABPs and RDLG undergo anisotropic phase separation (Fig.~\ref{fig_micro_phasediagram}).
In this section, we investigate the properties of phase separation of uniaxial ABPs in more detail.
We focus on the nucleation of persistent gas bubbles and the possibility of microphase separation, which have been found in recent studies~\cite{Shi2020}.

% ---------------- figure ----------------
\begin{figure*}[t]
\centering
\includegraphics[scale=1]{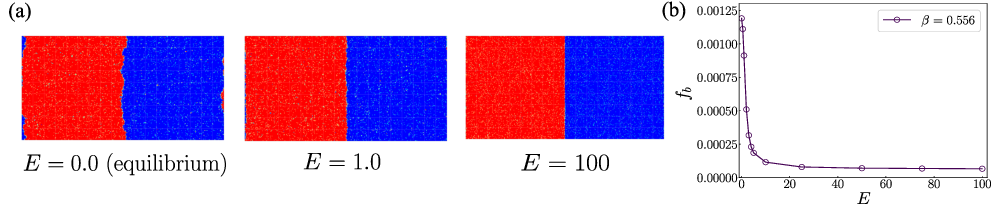}
\caption{Absence of gas bubbles in RDLG.
(a) Typical snapshots in the steady state for three values of $E$.
The colors represent the particle density from $0$ (blue) to $1$ (red). (b) Bubble fraction $f_b$ as a function of $E$.
In all figures, the parameters are chosen as $\rho=0.5$, $\beta=0.556$, and $(L_x,L_y)=(720,360)$.}
\label{fig_rdlg_phasediagram_app}
\end{figure*}
% ----------------------------------------

\subsection{Anisotropy-induced removal of gas bubbles}
\label{sec:anisotropy-induced_removal_of_gas_bubbles}
In Fig.~\ref{fig_micro_phasediagram_app}(a), we show typical density fields in the phase-separated states for three different values of $\epsilon$.
The detailed procedure for drawing this figure is given in Appendix~\ref{appendix:supplemental information of Fig5}.
From this figure, we find that for $\epsilon=0$, numerous gas bubbles are nucleated within the liquid phase.
Throughout this paper, we use a ``gas bubble'' to refer to a connected region of the gas phase surrounded by the largest liquid phase.
Note that we regard the largest gas phase as the gas reservoir and not as the gas bubble (see Appendix~\ref{appendix:numerical procedure to detect gas bubbles} for the method to detect gas bubbles).
As $\epsilon$ increases, the number of gas bubbles decreases.
For sufficiently large values of $\epsilon$ (e.g., $\epsilon=0.02$), the presence of gas bubbles becomes less evident.
To quantitatively characterize this observation, we define the bubble fraction as
\begin{eqnarray}
f_b := \frac{S_\mathrm{bubble}}{S},
\label{eq:def of bubble fraction}
\end{eqnarray}  
where $S:=L_xL_y$ and $S_\mathrm{bubble}$ is the total area occupied by gas bubbles.
We plot $f_b$ as a function of $\epsilon$ in Fig.~\ref{fig_micro_phasediagram_app}(b), which shows that the fraction of gas bubbles monotonically decreases as $\epsilon$ increases.
For sufficient large $\epsilon$, $f_b$ reaches zero, indicating the absence of gas bubbles.
This observation demonstrates that the uniaxial self-propulsion prevents the nucleations of gas bubbles.

In isotropic ABPs (i.e., $\epsilon=0$), the nucleation of gas bubbles has been examined in Ref.~\cite{Shi2020}, which has revealed a connection between the existence of gas bubbles and a novel type of phase separation called microphase separation~\cite{Tjhung2018}.
To briefly explain the previous results in Ref.~\cite{Shi2020}, we focus on the size distribution of gas bubbles divided by the total liquid area, $n(a)/S_\mathrm{liq}$, where $a$ is the area of a single bubble.
In Fig.~\ref{fig_micro_phasediagram_app}(d), we plot $n(a)/S_\mathrm{liq}$ for isotropic ABPs.
We find that $n(a)/S_\mathrm{liq}$ for large $a$ fits well with the power-law decay observed in the reduced bubble model~\cite{Shi2020}:
\begin{equation}
    \frac{n(a)}{S_{\rm liq}} \sim a^{\alpha} \ \ \ (\alpha = -1.77).
\end{equation}
Considering that the bubble fraction, $f_b$, and the size distribution, $n(a)$, are related as~\footnote{
In Ref.~\cite{Shi2020}, cutoff parameter $a_c$ is introduced as the upper limit of the integration in Eq.~\eqref{eq:relation between fb and n(a)}. In this study, we simply use $S_{\textrm{gas}}$ as the upper limit since we do not have a systematic procedure for determining $a_c$. The size of each bubble does not surpass the total gas area.}
\begin{eqnarray}
f_b = \frac{1}{S}\int_0^{S_\mathrm{gas}} an(a)da, 
\label{eq:relation between fb and n(a)}
\end{eqnarray}
we can derive the system size dependence of $f_b$ as
\begin{eqnarray}
f_b \sim \chi_\mathrm{liq} \chi_\mathrm{gas}^{\alpha+2} S^{\alpha+2}.
\label{eq:relation between fb and S}
\end{eqnarray}
Here, $\chi_\mathrm{liq}:=S_{\rm liq}/S$ and $\chi_\mathrm{gas}:= 1 - \chi_\mathrm{liq}$ represent the area fractions of the liquid and gas phases, respectively, and are nearly independent of the system size, $S$.
Thus, as $S$ increases, $f_b$ is expected to increase until it reaches the area fraction of the gas phase, $\chi_\mathrm{gas}$.
This implies that the whole gas phase exists as persistent gas bubbles surrounded by the liquid phase.
This state has been defined as the microphase-separated state~\cite{Shi2020}.

As seen in Fig.~\ref{fig_micro_phasediagram_app}(a), we find that gas bubbles are still observed for small but finite $\epsilon$.
We consider whether the size distribution of such gas bubbles can show the power-law decay as observed in isotropic ABPs (i.e., $\epsilon=0$).
In Fig.~\ref{fig_micro_phasediagram_app}(d), we plot $n(a)/S_\mathrm{liq}$ for $\epsilon=0.002$.
In contrast to the isotropic case, the bubble size distribution does not show the power-law behavior.
Note that this result is not attributed to the finite-size effect since $n(a)/S_\mathrm{liq}$ for different system sizes fall on a universal curve.
More specifically, $n(a)/S_\mathrm{liq}$ for $\epsilon=0.002$ decays faster than $a^{-2}$.
From Eq.~\eqref{eq:relation between fb and S}, $f_b$ is expected to converge to zero in the large system size limit, implying that uniaxial ABPs undergo macroscopic phase separation rather than microphase separation.
Thus, we confirm that the type of phase separation significantly changes by the anisotropic self-propulsion.
We also plot the $\epsilon$ dependence of $n(a)/S_\mathrm{liq}$ for $(L_x, L_y)=(2880, 1440)$ in Fig.~\ref{fig_micro_phasediagram_app}(e), which shows that the functional form of $n(a)/S_\mathrm{liq}$ is changed by a small amount of $\epsilon$.
This suggests that microphase separation can be prohibited even for extremely small $\epsilon$ (e.g., $\epsilon=0.0005$), though we need a more detailed finite-size scaling analysis to draw a conclusion.

We comment on possible gas bubbles in RDLG.
Note that previous studies on RDLG have not reported any possibility of microphase separation.
As shown in Fig.~\ref{fig_rdlg_phasediagram_app}(a), the nucleation of gas bubbles is hardly observed in typical snapshots for large systems, and macroscopic phase separation is expected to appear regardless of the strength of anisotropy.
The bubble fraction, $f_b$, plotted in Fig.~\ref{fig_rdlg_phasediagram_app}(b) suggests that the nucleation of gas bubbles is suppressed by anisotropic \red{driving force} $E$ in a similar way to uniaxial ABPs.

% ---------------- figure ----------------
\begin{figure}[t]
\centering
\includegraphics[scale=1]{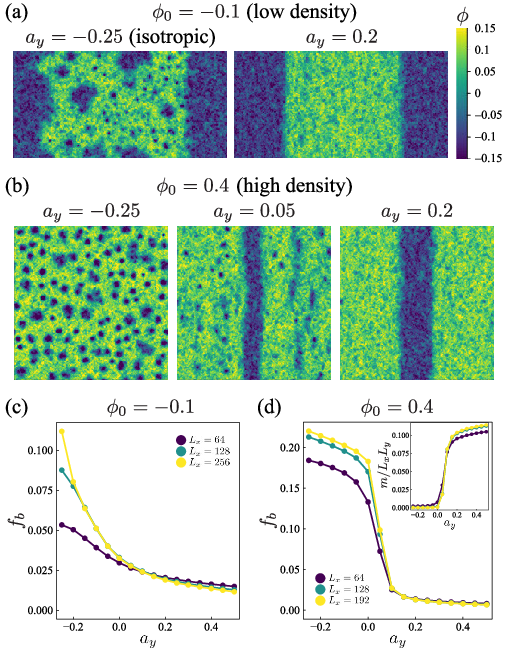}
\caption{Suppression of gas bubbles by anisotropy in the coarse-grained model.
For a low-density condition [$(\phi_0, \lambda, \zeta) = (-0.1, 0.5, 5)$], we show (a) typical snapshots for $(L_x, L_y) = (256, 128)$ and (c) bubble fraction $f_b$ as a function of anisotropy strength $a_y$ for three system lengths $L_x$ ($= 2L_y$).
For a high-density condition [$(\phi_0, \lambda, \zeta) = (0.4, 1, 4)$], we show (b) typical snapshots for $(L_x, L_y) = (192, 192)$ and (d) bubble fraction $f_b$ as in (c).
In the inset of (d), we plot the $a_y$ dependence of $m$, an order parameter for macroscopic phase separation.}
\label{Fig:cont}
\end{figure}
% ----------------------------------------

\subsection{Nonlinear coarse-grained model}
\label{sec:nonlinear_coarse-grained_model}
Though the linear coarse-grained model [Eq.~\eqref{Eq:cgmodel_linear}] succeeds in explaining the homogeneous state far from the critical point as discussed in Sec.~\ref{sec:homogeneous_state_properties}, it cannot describe phase separation since nonlinear terms are not included.
In previous studies on isotropic ABPs~\cite{Shi2020}, the qualitative features of microphase separation and the mechanism behind the observed persistent gas bubbles have been demonstrated using a coarse-grained model called Active Model B+ (AMB+)~\cite{Tjhung2018,Caballero2018}.
To discuss the observed suppression of gas bubbles by the anisotropic self-propulsion from a general perspective, we consider an anisotropic extension of AMB+:
\begin{align}
    \partial_t \phi = & a_x {\partial_x}^2 \phi + a_y {\partial_y}^2 \phi + \bm{\nabla}^2 (b \phi^3 - K \bm{\nabla}^2 \phi + K' \bm{\nabla}^4 \phi) \nonumber \\
    & + \lambda \bm{\nabla}^2 (\bm{\nabla} \phi)^2 - \zeta \bm{\nabla} \cdot [(\bm{\nabla}^2 \phi) \bm{\nabla} \phi] - \sqrt{2 D} \bm{\nabla} \cdot \bm{\xi},
    \label{Eq:cgmodel_nonlinear}
\end{align}
which is also regarded as a nonlinear extension (i.e., adding the $b$, $\lambda$, and $\zeta$ terms) of Eq.~\eqref{Eq:cgmodel_linear}.
The $b$ term can be derived from a coarse-grained Hamiltonian, and the $\lambda$ and $\zeta$ terms reflect the violation of the time-reversal symmetry~\cite{Tjhung2018}.
To improve numerical stability, the higher-order gradient term with a small $K'$ is also introduced.
This term is irrelevant in the RG sense (see Appendix~\ref{App:cgmodel_rg} for the detail) and is not expected to affect the qualitative phase behavior.
For simplicity, the effect of anisotropy is minimally retained in the difference between $a_x$ and $a_y$.

Throughout the numerical study of Eq.~\eqref{Eq:cgmodel_nonlinear}, we set $a_x = -0.25$, $b = 0.25$, $K = 1$, $K'=0.2$ and $D = 0.5$.
We take $(\phi_0, \lambda, \zeta) = (-0.1, 0.5, 5)$ and $(0.4, 1, 4)$ as low- and high-density cases, respectively, where $\phi_0$ is the spatial average of $\phi (\bm{r}, t)$.
The strength of anisotropy is controlled by $a_y$ ($\geq a_x$).
Considering the periodic boundary conditions along both axes [$\phi (x + L_x, y, t) = \phi (x, y + L_y, t) = \phi (x, y, t)$], we perform numerical integration of Eq.~\eqref{Eq:cgmodel_nonlinear} by the explicit Euler method (see Appendix~\ref{App:cgmodel_simulation} for the detail).
We regard the regions with $\phi < 0$ and $\phi > 0$ as the gas and liquid phases, respectively.

We explain the isotropic limit ($a_x = a_y$) with the present parameter set.
In the low-density case, we observe phase separation with persistent gas bubbles [Fig.~\ref{Fig:cont}(a), left], which is similar to the behavior of uniaxial ABPs [Fig.~\ref{fig_micro_phasediagram_app}(a), left].
In the high-density case, we observe microphase separation, where gas bubbles are present throughout the system [Fig.~\ref{Fig:cont}(b), left].
Such phase behaviors are consistent with the previous observations in the isotropic AMB+~\cite{Tjhung2018}.

We consider the effect of anisotropy on phase separation with gas bubbles [Fig.~\ref{Fig:cont}(a)].
Similarly to the observation in uniaxial ABPs [Fig.~\ref{fig_micro_phasediagram_app}(b)], we find the suppression of bubble fraction $f_b$ as shown in Fig.~\ref{Fig:cont}(c).
This suggests that the minimal extension of AMB+ (i.e., $a_x \neq a_y$) is sufficient to explain the qualitative behavior of uniaxial ABPs.
We next examine the effect of anisotropy on microphase separation [Fig.~\ref{Fig:cont}(b)].
We find that microphase separation discontinuously changes into macroscopic phase separation, indicated by the abrupt change in $f_b$ [Fig.~\ref{Fig:cont}(d)].
In addition, we define an order parameter for macroscopic phase separation along the $x$-axis as $m := S (k_x = 2 \pi / L_x, 0)$, where the structure factor is defined as $S(\bm{k}) := \braket{|\tilde{\phi} (\bm{k})|^2} / (L_x L_y)$ with $\tilde{\phi} (\bm{k}) := \int d^2 \bm{r} \, e^{- i \bm{k} \cdot \bm{r}} \phi (\bm{r})$.
As shown in the inset of Fig.~\ref{Fig:cont}(d), the discontinuous change in $m$ also suggests the discontinuous transition between microphase separation and macroscopic phase separation.

Let us focus on the case with $a_x < 0 < a_y$ [see the right panels of Figs.~\ref{Fig:cont}(a) and (b)] to consider why strong anisotropy suppresses gas bubbles and stabilizes macroscopic phase separation.
We neglect the noise term in Eq.~\eqref{Eq:cgmodel_nonlinear} by the mean-field approximation, which has been used in the previous studies~\cite{Solon2018a,Solon2018b,Tjhung2018}.
Then, the linearized equation for $\phi - \phi_0$ is obtained in the Fourier space as
\begin{equation}
    \partial_t \tilde{\phi} (\bm{k}, t) = -(a_x  {k_x}^2 + a_y {k_y}^2 + K |\bm{k}|^4 + K' |\bm{k}|^6) \tilde{\phi} (\bm{k}, t).
\end{equation}
From $a_x < 0 < a_y$, $K > 0$, and $K' > 0$, we see that the most unstable wavevector is along the $k_x$-axis.
Thus, we approximately neglect the modulation in the $y$ direction and replace Eq.~\eqref{Eq:cgmodel_nonlinear} by $\partial_t \phi = {\partial_x}^2 \mu$, where $\mu (x, t) := a_x \phi + b \phi^3 - K {\partial_x}^2 \phi + K' {\partial_x}^4 \phi + (\lambda - \zeta / 2) (\partial_x \phi)^2$.
Here, chemical potential $\mu$ is a local quantity, in contrast to the isotropic limit ($a_x = a_y$), where nonlocality of chemical potential can lead to phase separation with gas bubbles and microphase separation~\cite{Tjhung2018}.
Thus, macroscopic phase separation is expected to appear for $a_x < 0 < a_y$.

% ---------------- figure ----------------
\begin{figure*}[t]
\centering
\begin{center}
\includegraphics[scale=1]{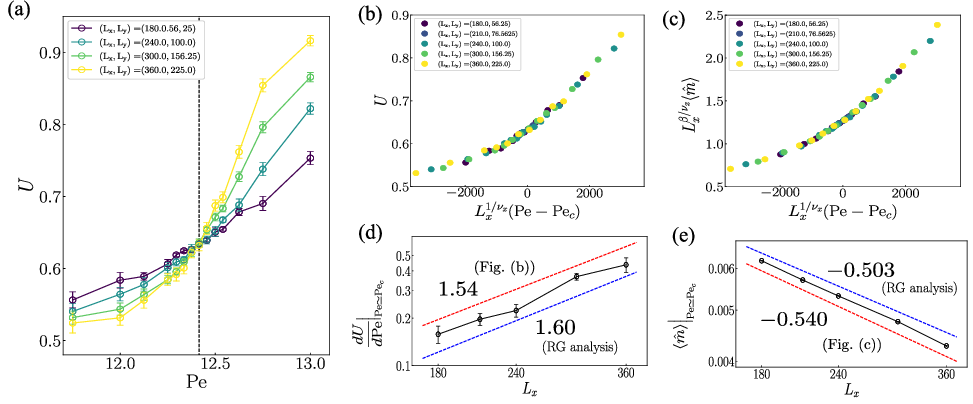} 
\end{center}
\caption{Finite-size scaling analysis for uniaxial ABPs. The parameters are chosen as $\rho=0.71$, $(\mu_{\parallel},\mu_{\perp},\mu_{\theta})=(1,0.25,1.5)$, and $\epsilon=0.01$.
(a) The Binder ratio $U$ as a function of $\textrm{Pe}$ for different system sizes.
(b) $U$ and (c) the rescaled order parameter $\braket{\hat{m}}$ as functions of the rescaled ${\rm Pe}$ with the best-fitted critical exponents ($\beta/\nu_x=0.540$, $1/\nu_x=1.54$).
(d) $\partial U / \partial \mathrm{Pe}$ and (e) $\langle \hat{m}\rangle$ against $L_x$ near the critical point in the log-log plot.
In (d) and (e), the red dashed lines represent $\partial U / \partial \mathrm{Pe} \propto {L_x}^{1 / \nu_x}$ and $\braket{\hat{m}} \propto {L_x}^{-\beta / \nu_x}$, respectively, with the critical exponents used in (b) and (c), and the blue dashed lines are counterparts for the expected universality class [Eq.~(\ref{eq:critical exponent of RG analysis})] based on the RG analysis.}
\label{fig_micro_criticalbehavior}
\end{figure*}
% ----------------------------------------

\section{Critical properties}
\label{sec:critical_phenomena}
Since uniaxial ABPs and RDLG share the common properties in the homogeneous and phase-separated states (see Secs.~\ref{sec:homogeneous_state_properties} and \ref{sec:phase_separation_properties}), we expect that the critical point for anisotropic phase separation in each model belongs to the same universality class.
In the following, we support this expectation using the RG analysis of the coarse-grained model [Eq.~\eqref{Eq:cgmodel_nonlinear}] and the finite-size scaling analysis of simulation data for uniaxial ABPs.

\subsection{Renormalization group analysis of coarse-grained model}
\label{subsec:RG}
We consider the critical phase transition between the homogeneous and phase-separated states in the coarse-grained model [Eq.~\eqref{Eq:cgmodel_nonlinear}] under sufficiently large anisotropy with $a_x < a_y$.
We first review the previous RG analyses of Eq.~\eqref{Eq:cgmodel_nonlinear} for $K' = \lambda = \zeta = 0$~\cite{Schmittmann1991,Schmittmann1993,Praestgaard1994,Praestgaard2000}.
Retaining only the relevant variables in the RG sense, we can obtain a model that is equivalent to a coarse-grained model of uniaxial dipolar ferromagnets, which have dipolar
long-range interactions~\cite{Schmittmann1991,Schmittmann1993,Praestgaard1994,Praestgaard2000} (see Appendix~\ref{App:dipolar_ferromagnet} for the detail).
At the two-loop level, the critical exponents for the coarse-grained model of uniaxial dipolar ferromagnets have been obtained~\cite{Schmittmann1993,Praestgaard1994,Praestgaard2000} as
\begin{eqnarray}
\beta = 0.315 \ , \ \nu_x = 0.626 \ \ \ (\text{Two-loop RG}).
\label{eq:critical exponent of RG analysis}
\end{eqnarray}
Here, $\beta$ is the exponent for the onset of the order parameter, and $\nu_x$ and $\nu_y \ (\simeq 2 \nu_x)$ are the exponents for the divergent correlation lengths along the $x$- and $y$-axes, respectively.
For RDLG, the finite-size scaling analysis of simulation data has been performed to obtain the critical exponents~\cite{Praestgaard2000} as
\begin{eqnarray}
\beta = 0.33(2) \ , \ \nu_x = 0.62(3) \ \ \ (\text{RDLG}).
\label{eq:critical exponent for RDLG}
\end{eqnarray}
These values coincide with the RG results [Eq.~\eqref{eq:critical exponent of RG analysis}] within the numerical error, suggesting that the critical point for anisotropic phase separation in RDLG belongs to the universality class of uniaxial dipolar ferromagnets.

Considering nonzero $\lambda$ and $\zeta$ to discuss the phase behavior of uniaxial ABPs (see Sec.~\ref{sec:phase_separation_properties}), we can show that $\lambda$ and $\zeta$ are irrelevant variables in the RG sense (see Appendix~\ref{App:cgmodel_rg} for the detail).
This suggests that the introduction of small $\lambda$ or $\zeta$ does not affect the critical properties of anisotropic phase separation, and the critical exponents remain the same as those given in Eq.~\eqref{eq:critical exponent of RG analysis}.
Thus, like RDLG, the critical point for anisotropic phase separation in uniaxial ABPs is expected to belong to the universality class of uniaxial dipolar ferromagnets.
Note that the irrelevance of $\lambda$ or $\zeta$ is further supported by the suppression of gas bubbles under strong anisotropy (see Fig.~\ref{Fig:cont}).

\subsection{Connection to uniaxial dipolar ferromagnets}
\label{sec:connection_to_uniaxial_dipolar_ferromagnets}
To study the critical point for anisotropic phase separation in uniaxial ABPs, we perform simulations with a fixed strength of anisotropy, $\epsilon = 0.01$.
Here, we assume that the critical exponents are not affected by the specific value of $\epsilon$.
First, assuming the law of rectilinear diameter~\cite{Watanabe2012,Garrabos2018}, we estimate the critical density as $\rho_c = 0.71$ (see Appendix~\ref{appendix:rough_estimation_of_critical_density} for the detail).
Next, we perform simulations with $\rho = \rho_c = 0.71$ to identify the universality class of the critical point using the anisotropic finite-size scaling analysis, which has been widely applied to critical phenomena in externally driven systems~\cite{Schmittmann1995,Leung1991,Wang1996, Nakano2021}.
Since the liquid and gas phases are separated along the $x$-axis for large $\mathrm{Pe}$ [Fig.~\ref{fig_micro_phasediagram}(a)], the degree of phase separation can be measured by an order parameter,
\begin{eqnarray}
\hat{m} := \frac{1}{L_xL_y} \sum_{j=1}^N e^{-i 2\pi x_j/L_x}.
\end{eqnarray}
The finite-size scaling hypotheses for $\langle \hat{m}\rangle$ and the Binder ratio, $U := \langle \hat{m}^2\rangle^2 / \langle \hat{m}^4\rangle$, are given as
\begin{eqnarray}
\langle \hat{m}\rangle = {L_x}^{-\beta/\nu_x}\mathcal{M}({L_x}^{1/\nu_x}\tau,L_y/{L_x}^{\nu_y/\nu_x};\epsilon,\rho)
\label{eq:anisotropic finite-size scaling hypotheses 1}
\end{eqnarray}
and
\begin{eqnarray}
U = \mathcal{U}({L_x}^{1/\nu_x}\tau,L_y/{L_x}^{\nu_y/\nu_x};\epsilon,\rho),
\label{eq:anisotropic finite-size scaling hypotheses 2}
\end{eqnarray}
respectively.
Here, $\tau := {\rm Pe}-{\rm Pe}_c$ is the distance from the critical point, and $\mathcal{M}$ and $\mathcal{U}$ are scaling functions.
Equations~\eqref{eq:anisotropic finite-size scaling hypotheses 1} and \eqref{eq:anisotropic finite-size scaling hypotheses 2} are extensions of the scaling hypotheses for isotropic systems with $\nu_x=\nu_y$~\cite{Nakano2021}, and the values of $\nu_x$ and $\nu_y$ can be different in anisotropic systems such as uniaxial ABPs and RDLG.
For $\nu_x \neq \nu_y$, to perform the finite-size scaling analysis, we need to vary the system size with $L_y/{L_x}^{\nu_y/\nu_x}$ fixed.
Though $\nu_y/\nu_x$ should be determined in principle by the finite-size scaling analysis, we choose $\nu_y/\nu_x = 2$, which has been commonly used for RDLG based on the RG analysis~\cite{Praestgaard1994,Praestgaard2000}.
Following this choice, we perform simulations with five different system sizes satisfying $L_y/{L_x}^2=1/24^2$: $(L_x, L_y) = (180,56.25)$, $(210,76.5625)$, $(240,100)$, $(300,156.25)$, and $(360,225)$.

The results of the finite-size scaling analysis are summarized in Fig.~\ref{fig_micro_criticalbehavior} (See Appendix~\ref{appendix:estimation_of_critical_exponents} for the detailed procedure).
Varying $\mathrm{Pe}$ from $11.5$ to $13.0$, we find that $U$ as a function of $\mathrm{Pe}$ for different system sizes approximately crosses at a unique point [Fig.~\ref{fig_micro_criticalbehavior}(a)], which suggests the presence of the critical point, $\mathrm{Pe}_c$.
By fitting $U(\tau,L_x)$ and $\langle \hat{m}\rangle(\tau,L_x)$ with second-order polynomials, we obtain $\mathrm{Pe}_c$ as
\begin{eqnarray}
\mathrm{Pe}_c=12.408(5)
\end{eqnarray}
and the critical exponents as
\begin{eqnarray}
\beta=0.35(4) \ , \ \nu_x=0.65(6) \ \ \ (\text{uniaxial ABPs}).
\label{eq:critical exponent for ABPs}
\end{eqnarray}
Using these obtained values, we find that the rescaled plots of $U$ and $\braket{\hat{m}}$ collapse onto universal curves [Figs.~\ref{fig_micro_criticalbehavior}(b) and (c)], which validates the anisotropic finite-size scaling hypotheses given by Eqs.~\eqref{eq:anisotropic finite-size scaling hypotheses 1} and \eqref{eq:anisotropic finite-size scaling hypotheses 2}.

The obtained $\beta$ and $\nu_x$ [Eq.~\eqref{eq:critical exponent for ABPs}] agree with the RG result for the coarse-grained model [Eq.~\eqref{eq:critical exponent of RG analysis}] and the simulation result of RDLG [Eq.~\eqref{eq:critical exponent for RDLG}] within the error margin.
This indicates that the critical phenomena in uniaxial ABPs belong to the universality class of uniaxial dipolar ferromagnets, as expected from the RG analysis (see Sec.~\ref{subsec:RG}).
To check the consistency of the obtained values of $\beta$ and $\nu_x$, we plot the $L_x$ dependence of $\partial U / \partial \mathrm{Pe}$ and $\braket{\hat{m}}$ at $\mathrm{Pe} = 12.415$ ($\simeq \mathrm{Pe}_c$) in Figs.~\ref{fig_micro_criticalbehavior}(d) and (e). According to Eqs.~\eqref{eq:anisotropic finite-size scaling hypotheses 1} and \eqref{eq:anisotropic finite-size scaling hypotheses 2}, the slopes of $\partial U / \partial \mathrm{Pe}$ and $\braket{\hat{m}}$ on the logarithmic scale are $1/\nu_x$ and $-\beta/\nu_x$, respectively.
Indeed, Figs.~\ref{fig_micro_criticalbehavior}(d) and (e) show that the slopes are comparable to the counterparts for the two-loop RG result [Eq.~\eqref{eq:critical exponent of RG analysis}].

% ---------------- table ----------------
\begin{table*}[t]
\begin{center}
\begin{tabularx}{185mm}{CCC}\hline
model & $\beta$ & $\nu_x$ \\ \hline\hline
2D uniaxial dipolar ferromagnet (two-loop RG) & 0.315 & 0.626\\
2D RDLG (Monte Carlo) & 0.33(2) & 0.62(3)\\
2D uniaxial ABPs (our study) & 0.35(4) & 0.65(6) \\
2D Ising model & 0.125 & 1 \\
3D Ising model (Monte Carlo) & 0.326 & 0.630\\ \hline\hline
\end{tabularx}
\caption{Comparison with the critical exponents of related models.} 
\label{tab_critical_exponents}
\end{center}
\end{table*}
% ---------------------------------------
\section{Discussion}
In this paper, to investigate the relation between MIPS and nonequilibrium phase separation caused by attractive interactions, we have studied the collective properties of 2D uniaxial ABPs, in which self-propulsion along the $x$-axis is favored.
Performing simulations, we have found three distinctive features of uniaxial ABPs: (i) generic long-range density correlation in the homogeneous state, (ii) anisotropic phase separation with suppressed nucleation of gas bubbles in contrast to isotropic ABPs, and (iii) critical phenomena that presumably belong to the universality class of 2D uniaxial ferromagnets with dipolar long-range interactions.
Since properties (i)-(iii) are common to RDLG, in which phase separation is induced by attractive interactions under external driving, we have established the connection between collective behaviors of uniaxial ABPs and RDLG.
Additionally, we have constructed a nonlinear coarse-grained model [Eq.~\eqref{Eq:cgmodel_nonlinear}] and substantiated the generality of properties (i)-(iii).

The critical exponents for the models related to this study are summarized in Table~\ref{tab_critical_exponents}, which points out that the critical behaviors of 2D uniaxial ABPs are close to those of the 3D Ising model rather than the 2D Ising model.
This property is consistent with the previous study concerning 2D uniaxial ferromagnets with dipolar long-range interactions~\cite{Aharony1973,Brezin1976}.
For 2D uniaxial dipolar ferromagnets, the effective increase in dimensionality has been attributed to the consequence of the long-range correlation caused by the dipolar interactions.
For 2D uniaxial ABPs, the long-range density correlation arising from the anisotropic nonequilibrium dynamics (see Sec.~\ref{sec:homogeneous_state_properties}) effectively increases the dimensionality from two to three, according to the analogy with uniaxial dipolar ferromagnets (see Appendix~\ref{App:dipolar_ferromagnet} for the detail).

Our results suggest that the origin of phase separation (i.e., self-propulsion or attractive interaction) is not essential for the collective behaviors of particles with anisotropic dynamics [Figs.~\ref{fig_schematic}(b) and (d)].
In contrast, for isotropic systems [Figs.~\ref{fig_schematic}(a) and (c)], the collective phenomena of self-propelled particles can be distinct from those of attractively interacting particles.
Specifically, in 2D isotropic ABPs, persistent gas bubbles or microphase separation can appear (see Sec.~\ref{sec:phase_separation_properties})~\cite{Tjhung2018,Shi2020}, and the universality class for critical phenomena can be different from the 2D Ising class~\cite{Caballero2018}.
Further studies are required to elucidate the condition for such differences in isotropic systems.

Recently, a wide range of active matter phases has been realized using biological~\cite{Szabo2006,Dunkel2013,Nishiguchi2017,Kawaguchi2017,Liu2019,Tan2022} and artificial~\cite{Deseigne2010,Palacci2013,Bricard2013,Kumar2014,Ginot2018,Deblais2018,Geyer2019,Chardac2021,Wang2021, Buttinoni2013} systems, especially under anisotropic conditions~\cite{Luo2022}.
The connection between uniaxial ABPs and RDLG suggests that active matter can serve as a platform for materializing the properties predicted for externally driven systems.
Though we have focused on uniaxial anisotropy in this study, it will be interesting to examine whether the collective behaviors of the standard DLG can be observed in ABPs with unidirectional anisotropy, which can be relevant to biological systems with chemical gradients.

\acknowledgements

We thank Kyogo Kawaguchi and Hiroshi Watanabe for the scientific discussions.
We also thank Yohsuke T. Fukai, Yoshihiro Michishita, and Yuki Nagai for their comments on programming and Rory Cerbus and Hiroshi Noguchi for their helpful comments.
The computations in this study were performed using the facilities of the Supercomputer Center at the Institute for Solid State Physics, the University of Tokyo.
This work was supported by JSPS KAKENHI Grant Numbers JP21J00034, JP22K13978 (to H.N.), and JP20K14435 (to K.A.).

\appendix
%%%%%%%%%%%%%%%%%%%%%%%%%%%%%%%%%%%%%%%%%%%%%%%%%%%%%%%%%%%%%%
% % ---------------- table ----------------
\begin{table*}[t]
\begin{center}
\begin{tabularx}{175mm}{CCCCC}\hline
model & direction of polarity & persistence time & intrinsic attractive force & anisotropy \\ \hline\hline
uniaxial ABPs & $0\sim 2\pi$ & finite & no & yes \\
RDLG  & $0$ or $\pi$ & zero & yes & yes \\
isotropic ABPs & $0\sim 2\pi$ & finite & no & no \\
equilibrium LG & not defined & zero limit & yes & no \\\hline\hline
\end{tabularx}
\caption{Basic features in microscopic implementation of uniaxial ABPs, RDLG, isotropic ABPs, and equilibrium LG.} 
\label{tab_microscopic_model}
\end{center}
\end{table*}
% % ---------------- table ----------------
\section{Comparison of microscopic dynamics between uniaxial ABPs and RDLG}
\label{App:diffandsimi_micro}
We compare the microscopic implementation of uniaxial ABPs and RDLG. In terms of single-particle dynamics, the fundamental aspects of the microscopic implementation are similar. Both models are based on overdamped dynamics, and the motion of particles is enhanced along the direction of polarity/\red{driving force}. However, by carefully comparing the microscopic implementation, we notice three distinct differences, which are summarized in Tab.~\ref{tab_microscopic_model}. They involve the direction of polarity, persistence time, and interparticle interaction: 
\begin{enumerate}
    \item Uniaxial ABPs allow a full 360-degree rotation of polarity, whereas RDLG restricts the angle of the driving field to either $\theta = 0$ or $\pi$.
    \item The persistence time of uniaxial ABPs is finite, similar to that of isotropic ABPs. In contrast, for RDLG, the direction of the \red{driving force} changes randomly, indicating that RDLG is characterized by the zero persistence time $\tau_p=0$
    \item RDLG contains both short-range attractive interaction and excluded volume interaction, whereas uniaxial ABPs involve only excluded volume interaction.
\end{enumerate}
Here, the persistence time $\tau_p$ is typically defined in ABPs as from correlation of the polarity $\bm{n}_i$. For example, in isotropic ABPs, the correlation of $\bm{n}_i$ is calculated as
\begin{align}
    \langle \bm{n}_i(s)\cdot \bm{n}_i(0) \rangle = e^{-\tau \mu_{\theta}s},
\end{align}
and consequently the persistence time is given by $\tau_p=1/\tau \mu_{\theta}$. In the comparison mentioned above, the concept of persistence time is extended to RDLG by considering the \red{driving force} as the equivalence of the polarity.

Similarly, we present the relationship between isotropic ABPs and equilibrium LG in Tab.~\ref{tab_microscopic_model}. Notably, the polarity is not defined in the equilibrium LG. However, an important point is that the passive Brownian particles, an off-lattice version of the equilibrium LG, correspond to the zero persistence time limit of isotropic ABPs. Then, by focusing on the aspects of the persistence time and intrinsic attractive force, we can interpret uniaxial ABPs and RDLG as one anisotropic extension of isotropic ABPs and equilibrium LG.

Due to the second point, RDLG cannot be linked to uniaxial ABPs with continuous changes of parameters such as the zero persistence time limit. To understand this point, we focus on the probability distribution of polarity angle, $P(\theta)$ for uniaxial ABPs, which is calculated as
\begin{eqnarray}
P(\theta) = \frac{1}{Z} \exp\Biggl(-\epsilon \frac{U(\theta)}{\tau \mu_{\theta}}\Biggr),
\label{eq:steady state of polarity}
\end{eqnarray}
where $Z$ is a normalization constant. To localize the polarity at $\theta = 0$ or $\pi$, we must change $\mu_{\theta}\to +0$ with $\epsilon$ fixed or $\epsilon \to +\infty$ with $\mu_{\theta}$ fixed. Clearly, under these continuous changes, the polarity cannot climb the potential barrier between $\theta = 0$ and $\theta = \pi$.

\section{Procedure for constructing phase diagram}
\label{App:procedure for constructing phase diagram}
In this Appendix, we explain the procedure for drawing the phase diagram [Figs.~\ref{fig_micro_phasediagram}(c) and (d)] in details.

As the initial state, we prepare the half-filling state by placing the particles in the right-half. After the relaxation run, we determine the high and low-density regions based on the fact that the center of the mass of the systems $x_{\mathrm{com}}$ coincides with the center of the high-density region. Specifically, we identify the high-density region as
\begin{eqnarray}
\Big[x_{\mathrm{com}}-\frac{L_x}{10},x_{\mathrm{com}}+\frac{L_x}{10}\Big] \times [0,L_y].
\end{eqnarray}
Also, from the fact that the center of the low-density region is the farthest from the center of the high-density region, we identify the low-density region as
\begin{eqnarray}
\Big[x_{\mathrm{com}}+\frac{L_x}{2}-\frac{L_x}{10},x_{\mathrm{com}}+\frac{L_x}{2}+\frac{L_x}{10}\Big] \times [0,L_y].
\end{eqnarray}
We then observe the density $\rho_l$ and $\rho_h$ in the high- and low-density regions. In the phase-separated state, the values of $\rho_l$ and $\rho_h$ give the coexisting (binodal) curve, which is drawn in Fig.~\ref{fig_micro_phasediagram}(c) and (d).

\section{Parameter details of Figs.~\ref{fig_micro_homobehavior} and \ref{fig_micro_homobehavior2}}
\label{App:paramdetail}
We set the simulation box to $L_x=L_y=360$. The particle number is set to $N=92016$ for uniaxial ABPs and $N=64800$ for RDLG, which respectively correspond to the density of $0.710$ and $0.50$. We start from the initial state in which the particles are randomly located with zero overlaps. We perform the relaxation run for $10^8$ time steps (i.e., time $=10^8 dt=2.0 \times 10^6$) for uniaxial ABPs and for $4.0\times 10^6$ Monte Carlo steps for RDLG. After that, we observe the structure factor $S(\bm{k})$. The real-space density correlation $\langle \rho(\bm{r}) \rho(\bm{0})\rangle$ is calculated by the inverse Fourier transformation of the structure factor $S(\bm{k})$.

We take the time average in the steady state and the ensemble average over different noise realizations. For uniaxial ABPs, the ensemble average is performed over $28$ different noise realizations, and the time average is performed over 400 samples obtained every $10^6$ time steps (i.e., time $=10^6 dt=20000$). For RDLG, the ensemble average is performed over $96$ different noise realizations, and the time average is performed over 400 samples obtained every $20000$ Monte Carlo steps.

\section{Relation to equilibrium uniaxial dipolar ferromagnet}
\label{App:dipolar_ferromagnet}
For RDLG, it is known that the specific patterns of structure factor $S(\bm{k})$ involving the long-range correlations are analogous to the long-range nature of the uniaxial dipolar system.
Here, we give the definition of uniaxial dipolar ferromagnet~\cite{Aharony1973} and briefly discuss the analogy between the density correlation of uniaxial ABPs and the spin correlation of uniaxial dipolar ferromagnet.

We start with the Heisenberg model with the short-range exchange interaction and long-range dipolar interaction. The Heisenberg spin $\bm{S}_{\bm{R}}$ is defined on the two-dimensional square lattice $\{\bm{R} = (n_x, n_y)\ |\ n_x,n_y=0,\pm1,\pm2,\cdots\}$, where the lattice constant is set to $1$. The Hamiltonian $\mathcal{H}$ of this model consists of the short-range exchange interaction and long-range dipolar interaction, which is expressed as
\begin{align}
\mathcal{H} =& - G \sum_{\bm{R} \neq \bm{R}'} \sum_{\alpha,\beta} \Biggl(-\frac{\delta_{\alpha \beta}}{|\bm{R}-\bm{R}'|^2}+\frac{(R_{\alpha}-R'_{\alpha})(R_{\beta}-R'_{\beta})}{|\bm{R}-\bm{R}'|^4}\Biggr) S^{\alpha}_{\bm{R}}S^{\beta}_{\bm{R}'} \nonumber \\
&- \frac{1}{2} J \sum_{\bm{R}} \sum_{\bm{\delta}}\bm{S}_{\bm{R}}\cdot \bm{S}_{\bm{R}+\bm{\delta}},
\end{align}
where $\sum_{\bm{R}}\sum_{\bm{\delta}}$ runs over all nearest-neighbor pairs.
Let us impose the uniaxial condition where the Heisenberg spin $\bm{S}_{\bm{R}}$ is restricted to pointing in the direction of the $y$-axis: $\bm{S}_{\bm{R}}=(0,S_{\bm{R}},0)$. The model reduces to the Ising model with anisotropic interaction:
\begin{align}
\mathcal{H} =& - G \sum_{\bm{R} \neq \bm{R}'} \Biggl(-\frac{1}{|\bm{R}-\bm{R}'|^2}+\frac{(R_{y}-R'_{y})^2}{|\bm{R}-\bm{R}'|^4}\Biggr) S_{\bm{R}}S_{\bm{R}'} \nonumber \\
&- \frac{1}{2} J \sum_{\bm{R}} \sum_{\bm{\delta}}S_{\bm{R}} S_{\bm{R}+\bm{\delta}} .
\end{align}
This model is called the uniaxial dipolar ferromagnet.

In the Fourier space, the dipolar part of the Hamiltonian is expanded near the $\bm{k}=\bm{0}$ as
\begin{multline}
- G \sum_{\bm{R} \neq \bm{R}'} \Biggl(-\frac{1}{|\bm{R}-\bm{R}'|^2}+\frac{(R_{y}-R'_{y})^2}{|\bm{R}-\bm{R}'|^4}\Biggr) \\ 
= a_1\Big(\frac{k_y}{k}\Big)^2 - a_2 k_y^2 - \big(a_3 + a_4\bm{k}^2 \big) + \cdots,
\end{multline}
where $\{ a_i \}_{i=1,\cdots,4}$ is a set of numerical constants depending on the lattice structure.
By expanding the short-range part of Hamiltonian in the same way, we rewrite the Hamiltonian as
\begin{align}
\mathcal{H} =& - \int \frac{d^2\bm{k}}{(2\pi)^2} \Biggl(r_0 + \bm{k}^2 - h_0k_y^2 + g_0 \frac{k_y^2}{\bm{k}^2}\Biggr) S_{\bm{k}}S_{\bm{k}'} \nonumber \\
&- u_0 \int \frac{d^2\bm{k}_1}{(2\pi)^2} \int \frac{d^2\bm{k}_2}{(2\pi)^2} \int \frac{d^2\bm{k}_3}{(2\pi)^2}S_{\bm{k}_1}S_{\bm{k}_2}S_{\bm{k}_3}S_{-\bm{k}_1-\bm{k}_2-\bm{k}_3},
\end{align}
where we ignore the higher-order terms in $S_{\bm{k}}$. The values of the numerical factor are given in Ref.~\cite{Aharony1973}.

The equilibrium state of this system is described by the canonical ensemble. In the disordered state, the linear approximation leads to the static spin-spin correlation:
\begin{eqnarray}
\langle S(\bm{k})S(\bm{k}')\rangle = C(\bm{k}) \delta(\bm{k}+\bm{k}')
\end{eqnarray}
with
\begin{eqnarray}
C(\bm{k}) = \frac{T \bm{k}^2}{r_0 k_x^2 + (r_0 + g_0)k_y^2 - h_0 k_y^2 \bm{k}^2 + \bm{k}^4}.
\end{eqnarray}
This form is the special case of Eq.~(\ref{Eq:strfac_linear}), indicating that uniaxial ABPs acquire dipolar-like long-range natures. As discussed in Appendix~\ref{App:cgmodel_rg}, this feature determines the universality class of critical phenomena.

\begin{figure*}[t]
\centering
\includegraphics[scale=1]{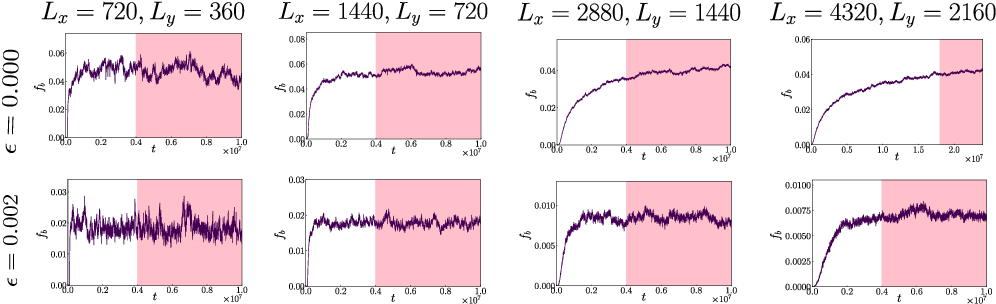}
\caption{Typical time dependence of the order parameter $\langle \hat{m}\rangle$, averaged over differet noise realizations. The parameters are the same as Fig.~\ref{fig_micro_phasediagram_app} [$\rho=0.765$, $(\mu_{\parallel},\mu_{\perp},\mu_{\theta})=(1.0,0.3125,2.75)$, and $\textrm{Pe} = 100.0$]. The observation run is performed within the red region.}
\label{SFig:micro_microbubbles_relaxation}
\end{figure*}
% ----------------------------------------
\section{Supplemental information of Fig.~\ref{fig_micro_phasediagram_app}}
\label{appendix:supplemental information of Fig5}
The microscopic simulation in the phase-separated phase of uniaxial ABPs is performed to examine the nucleation of bubbles, whose results are summarized in Fig.~\ref{fig_micro_phasediagram_app}. Here, we explain how to draw them.

\subsection{Relaxation run for observing gas bubbles}
\label{appendix:relaxation run for observing microbubbles}
The simulation box is a rectangle with the ratio of $L_x:L_y=2:1$. We prepare an initial configuration by placing the particles in the region $0 < x < L_x/2$ and perform the relaxation run. In Fig.~\ref{SFig:micro_microbubbles_relaxation}, we present the relaxation process of the different system sizes for $\epsilon=0$ and $0.002$, where the simulation data is averaged over $2\sim8$ different noise realizations. From this figure, we immediately notice that the relaxation time is significantly longer for the isotropic systems ($\epsilon=0$) compared to the anisotropic system. Additionally, the relaxation time increases as the system size becomes larger. According to this observation, we basically perform the relaxation run for $2.0 \times 10^8$ time steps (i.e., time $=2.0 \times 10^8 dt=4.0 \times 10^6$), and after that, perform the observation run for $3.0 \times 10^8$ time steps (i.e., time $=3.0 \times 10^8 dt=6.0 \times 10^6$). There is one exceptional case, specifically when $(L_x,L_y)=(4320,2160)$ with $\epsilon=0.000$, where the relaxation time is notably longer. In this specific case, we perform the relaxation run for $9.0 \times 10^8$ time steps (i.e., time $=9.0 \times 10^8 dt=18.0 \times 10^6$).

\subsection{Numerical procedure to detect gas bubbles}
\label{appendix:numerical procedure to detect gas bubbles}
After a sufficiently long relaxation run, we observe the bubble fraction, $f_b$, and the size distribution of gas bubbles, $n(a)$. For this observation, we divide the simulation box into square cells with a width of $\delta$, and calculate the density field as the collection of the local density. Figure~\ref{fig_micro_phasediagram_app}(a) draws the density field obtained using a bin size of $\delta=2.0$. 

The liquid and gas phases are distinguished based on the local density $\rho(\bm{r})$. The gas phase is designated by $\rho(\bm{r})<0.765$, while the liquid phase is designated by $\rho(\bm{r})>0.765$. As mentioned in the main text, the gas bubble is defined as the connected region of gas inside the liquid phase. The largest gas bubble is regarded as the gas reservoir and not as the gas bubble. To detect the connected regions of the gas phase, we used a Julia package JuliaImages.jl, which identifies the region connected to each other along the $x$ or $y$-axis. The bubble fraction $f_b$ is defined by Eq.~(\ref{eq:def of bubble fraction}). In the numerical procedure, we calculate the bubble fraction $f_b$ by rewriting Eq.~(\ref{eq:def of bubble fraction}) as
\begin{align}
f_b = \frac{S_\mathrm{gas} - a_\mathrm{max}}{S},
\end{align}
where $S_\mathrm{gas}$ is the total area of the gas phase and $a_\mathrm{max}$ is the maximum area of the gas phase (i.e. the area of the gas reservoir).

In Fig.~\ref{fig_micro_phasediagram_app}(b)-(e), to calculate $f_b$ and $n(a)/S_{\mathrm{liq}}$, we use the density field obtained with the bins $\delta=2.0$ for $(L_x,L_y)=(720,360)$ and $(L_x,L_y)=(1440,720)$, $\delta=4.0$ for $(L_x,L_y)=(2880,1440)$, and $\delta=6.0$ for $(L_x,L_y)=(4320,2160)$. We take the time average over $3000$ samples obtained every $10^5$ time steps (i.e., time $=10^5 dt=2000$) and the ensemble average over $2-8$ noise realizations.

\section{Supplemental information of Fig.~\ref{fig_micro_criticalbehavior}}
The simulations near criticality are performed to examine the universality class, whose results are summarized in Fig.~\ref{fig_micro_criticalbehavior} of the main text. In this appendix, we elaborate on the procedure for obtaining the critical properties such as the position of the critical point and the universality class.

% ---------------- Table ----------------

\begin{table}[b]
\begin{center}
\begin{tabularx}{60mm}{CCC}\hline
$\mathrm{Pe}$ & $(480,400)$ & $(720,900)$ \\ \hline\hline
10.0 & 0.764 & 0.763\\
11.0 & 0.760 & 0.765 \\
12.0 & 0.776 & 0.771 \\
13.0 & 0.723 & 0.708 \\
14.0 & 0.713 & 0.702 \\
15.0 & 0.700 & 0.701\\\hline\hline
\end{tabularx}
\caption{Rectilinear diameter $(\rho_l+\rho_g)/2$ near the critical point for $(L_x,L_y)=(480,400)$ and $(720,900)$.} 
\label{tab_micro_critical_1}
\end{center}
\end{table}
% ---------------- Table ----------------

% ---------------- figure ----------------
\begin{figure*}[t]
\centering
\includegraphics[scale=1]{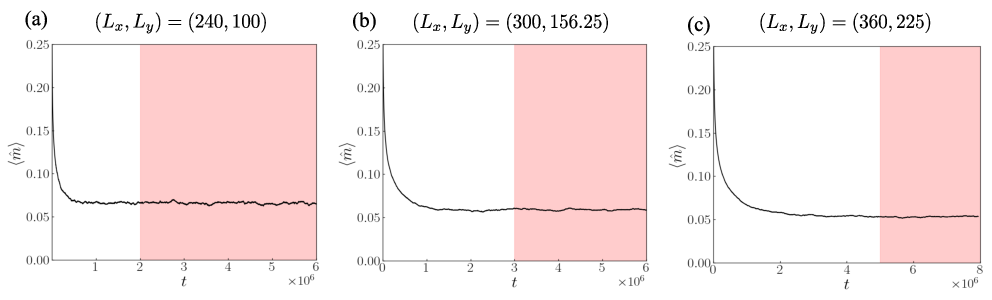}
\caption{Typical time dependence of the order parameter $\langle \hat{m}\rangle$, averaged over differet noise realizations. The parameters are $\rho=0.765$, $(\mu_{\parallel},\mu_{\perp},\mu_{\theta})=(1.0,0.3125,2.75)$, and $\textrm{Pe} = 12.415$. }
\label{SFig:micro_criticalbehavior_relaxation}
\end{figure*}
% ---------------- figure ----------------

\subsection{Rough estimation of critical density}
\label{appendix:rough_estimation_of_critical_density}
We first estimate the critical density by calculating the rectilinear diameter $(\rho_l+\rho_g)/2$ for various $\mathrm{Pe}$, where $\rho_l$ and $\rho_g$ are the densities in the liquid and gas phases, respectively.
Here, we summarize the supplemental information of this simulation.

The system size is set to $(L_x,L_y)=(480,400)$ and $(720,900)$. The density is set to $\rho=0.765$, which corresponds to the particle numbers $N=146 880$ and $495720$, respectively. We prepare an initial configuration by placing the particles in the region $0 < x < 200$ for $(L_x, L_y)=(480,400)$ and $0 < x < 300$ for $(L_x, L_y)=(720,900)$. In all simulations, we perform the relaxation run for $7.5 \times 10^7$ time steps (i.e., time $=7.5 \times 10^7 dt=1.5 \times 10^6$), and the observation run for $2.5 \times 10^7$ time steps (i.e., time $=2.5 \times 10^7 dt=0.5 \times 10^6$).

The simulation result is presented in Tab~\ref{tab_micro_critical_1}. In the large system size limit, the rectilinear diameter in the homogeneous state is equal to the global density of $0.765$, while at the critical point, it coincides with critical density $\rho_c$. From Tab.~\ref{tab_micro_critical_1}, we observe a distinct change in the rectilinear diameter between $\mathrm{Pe}=12.0$ and $\mathrm{Pe}=13.0$. Specifically, at $\mathrm{Pe}=12.0$, it closely matches the expected value of $0.765$, whereas at $\mathrm{Pe}=13.0$ it significantly deviates from this value. Based on this observation, we can infer that the critical P{\'e}clet number, $\mathrm{Pe}_c$, lies between $12.0 < \mathrm{Pe}_c < 13.0$ and the critical density, $\rho_c$, is estimated as $\approx 0.708$.

\subsection{Estimation of critical exponents}
\label{appendix:estimation_of_critical_exponents}
Based on the estimation of the critical density in the previous section, we set the density to $\rho=0.710$ and change the P{\'e}clet number, $\mathrm{Pe}$, from $\mathrm{Pe}=11.5$ to $\mathrm{Pe}=13.0$. As explained in main text, we set the system sizes to $(L_x,L_y) = (180,56.25)$, $(210,76.5625)$, $(240,100)$, $(300,156.25)$, and $(360,225)$. We show the typical time evolution of the ensemble average of the order parameter for $(L_x,L_y) = (240,100)$, $(300,156.25)$, and $(360,225)$ in Fig.~\ref{SFig:micro_criticalbehavior_relaxation}. This figure confirms that our simulation achieves the steady state after a sufficiently long relaxation run. Using the data within the red region, we take the time and ensemble averages for the order parameter $\langle \hat{m}\rangle$ and the Binder Parameter $U := \langle \hat{m}^2\rangle^2 / \langle \hat{m}^4\rangle$. The ensemble average is taken over $800$ different noise realizations for $(L_x,L_y) = (300,156.25)$ and $(360,225)$, and $500$ different noise realizations for $(L_x,L_y) = (180,56.25)$, $(210,76.5625)$, and $(240,100)$. The time average is performed by using the data every $4.0 \times 10^6$ time steps (i.e., time $=4.0 \times 10^6 dt=8.0 \times 10^4$) for $(L_x,L_y) = (360,225.5)$, $2.0 \times 10^6$ time steps (i.e., time $=2.0 \times 10^6 dt=4.0 \times 10^4$) for $(L_x,L_y) = (300,156.25)$, $1.0 \times 10^6$ time steps (i.e., time $=1.0 \times 10^6 dt=2.0 \times 10^4$) for $(L_x,L_y) = (240,100)$, $0.5 \times 10^6$ time steps (i.e., time $=0.5 \times 10^6 dt=1.0 \times 10^4$) for $(L_x,L_y) = (210,76.5625)$, and $0.25 \times 10^6$ time steps (i.e., time $=0.25 \times 10^6 dt=0.5 \times 10^4$) for $(L_x,L_y) = (180,56.25)$.

To estimate the critical exponents, we use the anisotropic finite-size scaling hypothesis Eqs.~(\ref{eq:anisotropic finite-size scaling hypotheses 1}) and (\ref{eq:anisotropic finite-size scaling hypotheses 1}). We refer to Ref.~\cite{Nakano2021} for a more detailed discussion of the anisotropic finite-size scaling. Since the scaling functions $\mathcal{M}$ and $\mathcal{U}$ are analytic, we can expand $\langle \hat{m}\rangle(\tau,L_x)$ and $U(\tau,L_x)$ around $\tau=0$ as
\begin{eqnarray}
\langle \hat{m}\rangle(\tau,L_x) = \sum_{n=0}^{\infty} \frac{\partial^n \mathcal{M}}{\partial \tau^n}\Big|_{\tau=0} (L_y/{L_x}^{\nu_y/\nu_x};\epsilon,\rho) {L_x}^{(n-\beta)/\nu_x} \tau^n,
\end{eqnarray}
\begin{eqnarray}
U(\tau,L_x) = \sum_{n=0}^{\infty} \frac{\partial^n \mathcal{U}}{\partial \tau^n}\Big|_{\tau=0} (L_y/{L_x}^{\nu_y/\nu_x};\epsilon,\rho) {L_x}^{n/\nu_x} \tau^n.
\end{eqnarray}
According to these expansions, we fit the simulation data $\langle \hat{m}\rangle$ and $U$ to the second-order polynomials to obtain the critical point $\mathrm{Pe}_c=12.408(5)$ and the critical exponents $\beta=0.35(4)$ and $\nu_x=0.65(6)$. For this fitting, the data within $-2000.0<{L_x}^{1/0.65}(\mathrm{Pe}-12.408)<2000.0$ are used.

% ---------------- figure ----------------
\begin{figure}[t]
\centering
\includegraphics[scale=1]{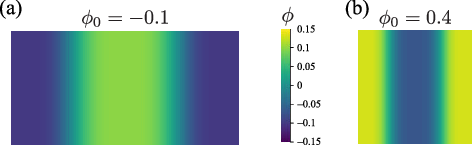}
\caption{Initial states used in simulations of the coarse-grained model.
    We show the initial states for (a) $\phi_0 = -0.1$ with $(L_x, L_y) = (256, 128)$ and (b) $\phi_0 = 0.4$ with $(L_x, L_y) = (192, 192)$, which correspond to Figs.~\ref{Fig:cont}(a) and (b), respectively.}
\label{SFig:cont_initstate}
\end{figure}

\begin{figure*}[t]
\centering
\includegraphics[scale=1]{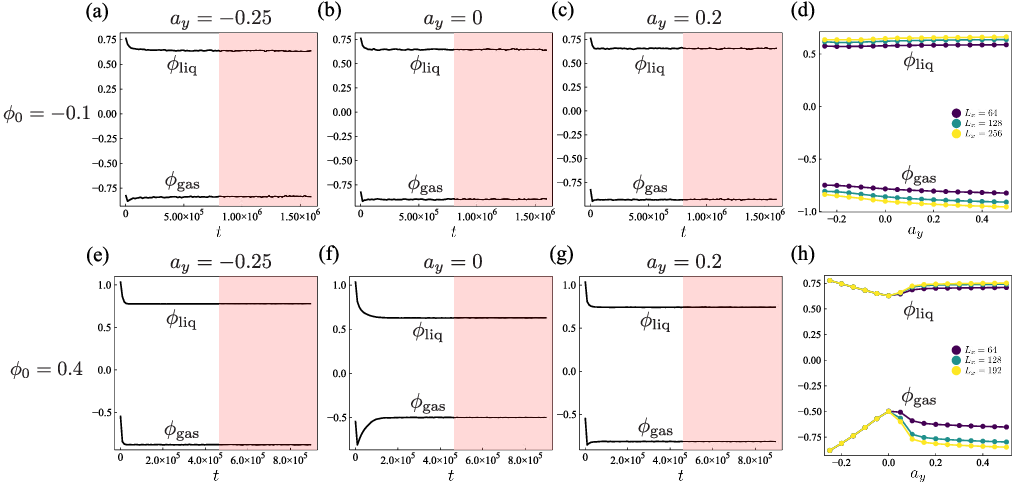}
\caption{Typical time and $a_y$ dependence of $\phi$ in the liquid and gas phases ($\phi_\mathrm{liq}$ and $\phi_\mathrm{gas}$, respectively), averaged over space and independent samples.
    We show the time evolution with (a) $a_y = -0.25$, (b) $a_y = 0$, and (c) $a_y = 0.2$ for $\phi_0 = -0.1$ and $(L_x, L_y) = (256, 128)$, as well as (e-g) the counterparts for $\phi_0 = 0.4$ and $(L_x, L_y) = (192, 192)$.
    The values at equally spaced 51 time points within the red region in (a-c) and (e-g) are used in time averaging to obtain the $a_y$ dependence of $\phi_\mathrm{liq}$ and $\phi_\mathrm{gas}$, which is plotted in (d) and (h).}
\label{SFig:cont_phi}
\end{figure*}
% ----------------------------------------

\section{Coarse-grained model}

\subsection{Numerical simulation}
\label{App:cgmodel_simulation}

For simulations of the coarse-grained model [Eq.~\eqref{Eq:cgmodel_nonlinear}],
\begin{align}
    \partial_t \phi = & a_x {\partial_x}^2 \phi + a_y {\partial_y}^2 \phi + \bm{\nabla}^2 (b \phi^3 - K \bm{\nabla}^2 \phi + K' \bm{\nabla}^4 \phi) \nonumber \\
    & + \lambda \bm{\nabla}^2 (\bm{\nabla} \phi)^2 - \zeta \bm{\nabla} \cdot [(\bm{\nabla}^2 \phi) \bm{\nabla} \phi] - \sqrt{2 D} \bm{\nabla} \cdot \bm{\xi},
    \label{SEq:aniso_ambp}
\end{align}
with $\braket{\xi_a (\bm{r}, t)} = 0$ and $\braket{\xi_a (\bm{r}, t) \xi_b (\bm{r'}, t')} = \delta_{ab} \delta (\bm{r} - \bm{r}') \delta (t - t')$, we discretize time as $t = n \Delta t$ and spatial coordinates as $x = i \Delta x$ and $y = j \Delta y$ with periodic boundary conditions.
Accordingly, we replace $\phi (x, y, t)$ by $\phi_{i,j}^n$ and $\xi_a (x, y, t)$ by $(\Delta x \Delta y \Delta t)^{-1/2} \xi_{a,i,j}^n$, where $\xi_{a,i,j}^n$ is a  Gaussian noise with $\braket{\xi_{a,i,j}^n} = 0$ and $\braket{\xi_{a,i,j}^n \xi_{b,i',j'}^{n'}} = \delta_{ab} \delta_{i i'} \delta_{j j'} \delta_{n n'}$.
Using the explicit Euler method, we replace Eq.~\eqref{SEq:aniso_ambp} by
\begin{equation}
    \phi_{i,j}^{n + 1} = \phi_{i,j}^n + [F (\phi)]_{i,j}^n \Delta t,
\end{equation}
where $[F (\phi)]_{i,j}^n$ is the discretized form of the right-hand side of Eq.~\eqref{SEq:aniso_ambp}.
To determine $[F (\phi)]_{i,j}^n$, we use the second-order central finite difference for the differential operators that appear in Eq.~\eqref{SEq:aniso_ambp} (i.e., $\partial_x$, $\partial_y$, ${\partial_x}^2$, and ${\partial_y}^2$), such as $[\partial_x f]_{i,j}^n = (f_{i+1,j}^n - f_{i-1,j}^n) / (2 \Delta x)$ and $[{\partial_x}^2 f]_{i,j}^n = (f_{i+1,j}^n - 2f_{i,j}^n + f_{i-1,j}^n) / \Delta x^2$.
The discretization parameters are chosen as $\Delta t = 0.01$ and $\Delta x = \Delta y = 1$, and the model parameters are fixed as $a_x = -0.25$, $b = 0.25$, $K = 1$, $K' = 0.2$, and $D = 0.5$ throughout the numerical study.
The other parameters are $(\lambda, \zeta) = (0.5, 5)$ for $\phi_0 = -0.1$ (low-density case) and $(\lambda, \zeta) = (1, 4)$ for $\phi_0 = 0.4$ (high-density case), where $\phi_0$ is the spatial average of $\phi (\bm{r}, t)$.
As the initial state for all the simulations, we use a phase-separated state, $\phi_\mathrm{init} (\bm{r}) := -2 \mathrm{sgn}(\phi_0) \exp[-(x - L_x / 2)^4 / (L_x / 4)^4] - C$, where $C$ is a constant to set the spatial average of $\phi_\mathrm{init} (\bm{r})$ to $\phi_0$ (Fig.~\ref{SFig:cont_initstate}).

We define the liquid and gas phases as the spatial regions satisfying $\phi (\bm{r}) > 0$ and $\phi (\bm{r}) < 0$, respectively.
In the same way as applied to uniaxial ABPs (see Appendix~\ref{appendix:numerical procedure to detect gas bubbles}), a Julia package (JuliaImages.jl) is used to detect the connected regions of the gas phase.
The size of each gas phase, $a$, is defined as the area of the regions that satisfy $\phi (\bm{r}) < 0$ and are connected to each other along the $x$ or $y$-axis.
The bubble fraction, $f_b$, which is plotted in Figs.~\ref{Fig:cont}(c) and (d), is calculated as $f_b := \braket{S_\mathrm{gas} - a_\mathrm{max}} / (L_x L_y)$, where $S_\mathrm{gas}$ and $a_\mathrm{max}$ are the total and maximum areas of the gas phase, respectively, and $\braket{\cdots}$ means the average over samples.

% ---------------- figure ----------------
\begin{figure*}[t]
\centering
\includegraphics[scale=1]{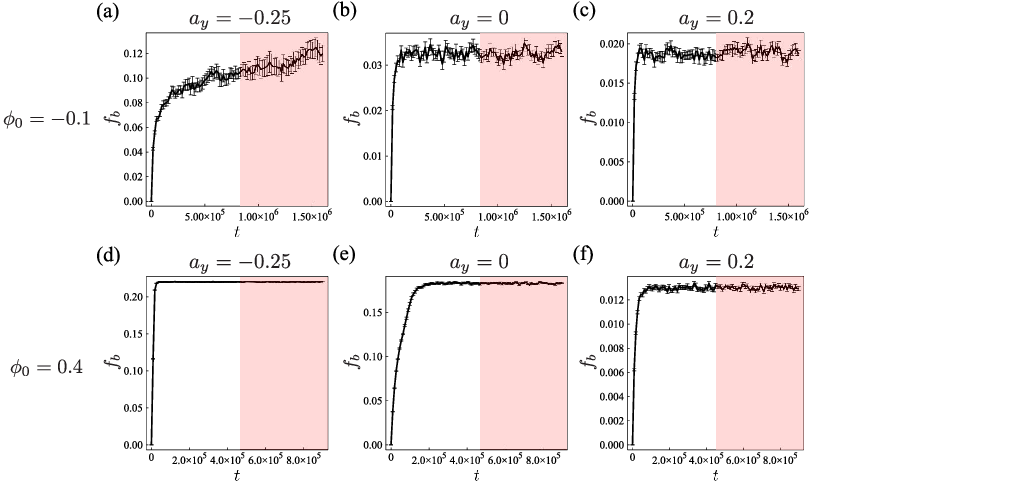}
\caption{Typical time dependence of bubble fraction $f_b$, averaged over independent samples.
    The same parameters as used in Figs.~\ref{SFig:cont_phi}(a-c) and (e-g) are used, and the error bar represents the standard error.
    The values at equally spaced 51 time points within the red region are used in time averaging to obtain the $a_y$ dependence of $f_b$, which is plotted in Figs.~\ref{Fig:cont}(c) and (d).}
\label{SFig:cont_fb}
\end{figure*}
% ----------------------------------------

To characterize the steady state using bubble fraction $f_b$ and order parameter $m$, independent samples are taken with different noise realizations.
For the low-density condition with $\phi_0 = -0.1$, which is used for Figs.~\ref{Fig:cont}(a) and (c), 1152 independent samples are taken with $10^7$ time steps (i.e., total time $= 10^7 \Delta t = 10^5$ for each sample) for $(L_x, L_y) = (64, 32)$, 1152 independent samples with $4 \times 10^7$ time steps for $(L_x, L_y) = (128, 64)$, and 24 independent samples with $1.6 \times 10^8$ time steps for $(L_x, L_y) = (256, 128)$.
For the high-density condition with $\phi_0 = 0.4$, which is used for Figs.~\ref{Fig:cont}(b) and (d), 1152 independent samples are taken with $10^7$ time steps for $(L_x, L_y) = (64, 64)$, 288 independent samples with $4 \times 10^7$ time steps for $(L_x, L_y) = (128, 128)$, and 128 independent samples with $9 \times 10^7$ time steps for $(L_x, L_y) = (192, 192)$.
To obtain the expectation values, we take the average over independent samples as well as the time average over 51 points in the last half of the total time.

We show the typical time evolution of $\phi$ in the liquid and gas phases ($\phi_\mathrm{liq}$ and $\phi_\mathrm{gas}$, respectively), averaged over space and independent samples [Figs.~\ref{SFig:cont_phi}(a-c) and (e-g)].
The points in the red region in Figs.~\ref{SFig:cont_phi}(a-c) and (e-g) are used in time averaging to obtain the $a_y$ dependence of $\phi_\mathrm{liq}$ and $\phi_\mathrm{gas}$, which is plotted in Figs.~\ref{SFig:cont_phi}(d) and (h).
Similarly, we show the typical time evolution of $f_b$ in Fig.~\ref{SFig:cont_fb}, in which the points in the red region are used to obtain the $a_y$ dependence of $f_b$ [Figs.~\ref{Fig:cont}(c) and (d)].
Note that, near the isotropic limit [$a_x = a_y$ ($= -0.25$)] for the low-density case ($\phi_0 = - 0.1$), the relaxation is slow as seen from Fig.~\ref{SFig:cont_fb}(a), and thus the values of $f_b$ plotted in Fig.~\ref{Fig:cont}(c) can be underestimated around $a_y = -0.25$, which is not essential for the current study.

\subsection{Renormalization group analysis}
\label{App:cgmodel_rg}

Assuming anisotropic systems with $a_y > 0$, we consider the critical phase transition between the homogeneous state and anisotropic phase separation that occurs as $a_x$ is changed.
Applying the approach by Martin, Siggia, Rose, Janssen, and de Dominicis (MSRJD)~\cite{Martin1973,Janssen1976,Dominicis1976,Tauber2014} to Eq.~\eqref{SEq:aniso_ambp}, we can obtain the probability density for a dynamical path of configurations $\{ \phi (\bm{r}, t) \}_{t \in [0, T]}$ as
\begin{equation}
    P[\phi] = \int D (i \bar{\phi}) \exp (-S[\phi, \bar{\phi}]).
\end{equation}
Here, dynamical action $S[\phi, \bar{\phi}]$ is given as
\begin{align}
    & S[\phi, \bar{\phi}] = \int_0^T dt \int d^2 \bm{r} \, \{\bar{\phi} [\partial_t \phi - a_x {\partial_x}^2 \phi - a_y {\partial_y}^2 \phi - b_x {\partial_x}^2 \phi^3 \nonumber \\
    & - b_y {\partial_y}^2 \phi^3 + K_{xx} {\partial_x}^4 \phi + K_{xy} {\partial_x}^2 {\partial_y}^2 \phi + K_{yy} {\partial_y}^4 \phi - K_{xxx}' {\partial_x}^6 \phi \nonumber \\
    & - K_{xxy}' {\partial_x}^4 {\partial_y}^2 \phi- K_{xyy}' {\partial_x}^2 {\partial_y}^4 \phi- K_{yyy}' {\partial_y}^6 \phi - \lambda_{xx} {\partial_x}^2 (\partial_x \phi)^2 \nonumber \\
    & - \lambda_{xy} {\partial_x}^2 (\partial_y \phi)^2 - \lambda_{yx} {\partial_y}^2 (\partial_x \phi)^2 - \lambda_{yy} {\partial_y}^2 (\partial_y \phi)^2 \nonumber \\
    & + \zeta_{xy} \partial_x ({\partial_y}^2 \phi \, \partial_x \phi) + \zeta_{yx} \partial_y ({\partial_x}^2 \phi \, \partial_y \phi) ] + D_{x} \bar{\phi} {\partial_x}^2 \bar{\phi} + D_{y} \bar{\phi} {\partial_y}^2 \bar{\phi}\},
    \label{SEq:action}
\end{align}
where we generalize the coupling constants, which are related to the original ones as $b_x = b_y = b$, $K_{xx} = K_{yy} = K$, $K_{xy} = 2 K$, $K'_{xxx} = K'_{yyy} = K'$, $K'_{xxy} = K'_{xyy} = 3 K'$, $\lambda_{xx} = \lambda_{yy} = \lambda - \zeta / 2$, $\lambda_{xy} = \lambda_{yx} = \lambda$, $\zeta_{xy} = \zeta_{yx} = \zeta$, and $D_x = D_y = D$.

Considering the tree-level renormalization group analysis of Eq.~\eqref{SEq:action}, we perform the scale transformation as $x \to c^{-1} x$ ($c > 1$).
Requiring the invariance of $a_y$, $K_{xx}$, and $D_x$ to consider the criticality of anisotropic phase separation, we can obtain the scaling of the other quantities: $y \to c^{-2} y$, $t \to c^{-4} t$, $\phi \to c^{1/2} \phi$, $\bar{\phi} \to c^{5/2} \bar{\phi}$, $a_x \to c^2 a_x$, $b_x \to c b_x$, $b_y \to c^{-1} b_y$, $K_{xy} \to c^{-2} K_{xy}$, $K_{yy} \to c^{-4} K_{yy}$, $K'_{xxx} \to c^{-2} K'_{xxx}$, $K'_{xxy} \to c^{-4} K'_{xxy}$, $K'_{xyy} \to c^{-6} K'_{xyy}$, $K'_{yyy} \to c^{-8} K'_{yyy}$, $\lambda_{xx} \to c^{-1/2} \lambda_{xx}$, $\lambda_{xy} \to c^{-5/2} \lambda_{xy}$, $\lambda_{yx} \to c^{-5/2} \lambda_{yx}$, $\lambda_{yy} \to c^{-9/2} \lambda_{yy}$, $\zeta_{xy} \to c^{-5/2} \zeta_{xy}$, $\zeta_{yx} \to c^{-5/2} \zeta_{yx}$, and $D_y \to c^{-2} D_y$.
Thus, $a_x$ and $b_x$ are relevant variables, the former of which works as a control parameter for the critical phase transition.
The other coupling constants, especially $K'$, $\lambda$, and $\zeta$, are irrelevant variables.

Neglecting all the irrelevant variables, we can obtain the effective action for the critical dynamics of Eq.~\eqref{SEq:aniso_ambp}:
\begin{align}
    S_\mathrm{eff}[\phi, \bar{\phi}] = \int_0^T dt \int d^2 \bm{r} [\bar{\phi} (& \partial_t \phi - a_x {\partial_x}^2 \phi - b_x {\partial_x}^2 \phi^3 \nonumber \\
    & + K_{xx} {\partial_x}^4 \phi) + D_{x} \bar{\phi} {\partial_x}^2 \bar{\phi}],
\end{align}
which coincides with the effective action for the randomly driven lattice gas~\cite{Schmittmann1991,Schmittmann1993,Praestgaard1994,Praestgaard2000}.

\bibliography{ref.bib}

\end{document}